\documentclass[sigconf,nonacm]{acmart}

\usepackage{amsmath,amsfonts}
\usepackage{algorithmic}
\usepackage{graphicx}
\usepackage{textcomp}
\usepackage{xcolor}
\usepackage{weiwAlgorithm}
\usepackage{listings}
\usepackage{enumitem}
\usepackage{amsthm}
\usepackage{titlesec}
\usepackage{subcaption}
\usepackage{array}
\usepackage{float}

\titleformat{\subsubsection}[block]
  {\normalfont\bfseries\color{black}}   
  {\thesubsubsection}{1em}{}            

\setcopyright{acmlicensed}
\copyrightyear{2018}
\acmYear{2018}
\acmDOI{XXXXXXX.XXXXXXX}

\acmConference[Sigmod '25]{International Conference on Management of Data}{June 22--27,
  2025}{Berlin, Germany}

\setlist[itemize]{leftmargin=1.2em, labelsep=0.5em}

\lstdefinelanguage{gopt}
{
  morekeywords={
    getV,
    expandE,
    patternStart,
    patternEnd,
    UNION,
    MATCH,
    RETURN,
    WHERE,
    AND,
    OR,
    SET,
    MERGE,
    ON,
    IN,
    class,
    implements,
    extends,
    interface,
    return,
    override,
    Double,
    join,
    select,
    group,
    order,
    new
  },
  morecomment=[l]{//},
}

\definecolor{eclipseBlue}{RGB}{42,0.0,255}
\definecolor{eclipseGreen}{RGB}{63,127,95}
\definecolor{eclipsePurple}{RGB}{127,0,85}

\begin{document}

\title{A Graph-native Optimization Framework for Complex Graph Queries}

\author{Bingqing Lyu, Xiaoli Zhou, Longbin Lai, Yufan Yang, Yunkai Lou, Wenyuan Yu, Ying Zhang$^{\ddag}$, Jingren Zhou}
\authornote{Both authors contributed equally to this research.}
\email{{bingqing.lbq,yihe.zxl, longbin.lailb, xiaofan.yyf, louyunkai.lyk, wenyuan.ywy, jingren.zhou}@alibaba-inc.com}
\email{ying.zhang@zjgsu.edu.cn}
\affiliation{%
  \institution{Alibaba Group, $^{\ddag}$Zhejiang Gongshang University}
  \country{China}
}




\begin{abstract}
  This technical report extends the SIGMOD 2025 paper "A Modular Graph-Native Query Optimization Framework" by providing a comprehensive exposition of \gopt’s advanced technical mechanisms, implementation strategies, and extended evaluations. While the original paper introduced \gopt’s unified intermediate representation (\ir) and demonstrated its performance benefits, this report delves into the framework’s implementation depth: (1) the full specification of \gopt’s optimization rules; (2) a systematic treatment of semantic variations (e.g., homomorphism vs. edge-distinct matching) across query languages and their implications for optimization; (3) the design of GOpt’s Physical integration interface, enabling seamless integration with transactional (Neo4j) and distributed (GraphScope) backends via engine-specific operator customization; and (4) a detailed analysis of plan transformations for LDBC benchmark queries.

\end{abstract}

%
%




\newcommand{\stitle}[1]{\vspace{0.5ex}\noindent{\bf #1}}
\newcommand{\etitle}[1]{\vspace{0.5ex}\noindent{\em\underline{#1}}}
\newcommand{\eetitle}[1]{\vspace{0.5ex}\noindent{\em{#1}}}
\newcommand{\eat}[1]{}

\newcommand{\patrel}{{\kw{CGP}}}
\newcommand{\patrels}{{\kw{CGPs}}}
\newcommand{\bgp}{{\kw{pattern}}}
\newcommand{\bgps}{{\kw{patterns}}}
\newcommand{\cgp}{{\kw{CGP}}}
\newcommand{\cgps}{{\kw{CGPs}}}
\newcommand{\estimation}{{\kw{CardinalityEstimation}}}

\newcommand{\kw}[1]{\textsf{#1}\xspace}
\newcommand{\gs}{{\kw{GraphScope}}}
\newcommand{\gaia}{{\kw{Gaia}}}
\newcommand{\glogue}{\kw{GLogue}}
\newcommand{\gopt}{\kw{GOpt}}
\newcommand{\glogs}{\kw{GLogS}}
\newcommand{\cypherplanner}{\kw{CypherPlanner}}

\newcommand{\btype}{\kw{BasicType}}
\newcommand{\btypes}{\kw{BasicTypes}}
\newcommand{\utype}{\kw{UnionType}}
\newcommand{\utypes}{\kw{UnionTypes}}
\newcommand{\unionall}{\kw{AllType}}
\newcommand{\unionalls}{\kw{AllTypes}}
\newcommand{\xtype}{\kw{DynType}}
\newcommand{\elemtype}{\lambda}
\newcommand{\type}{\tau}
\newcommand{\prop}{\pi}

\newcommand{\graphtype}{\kw{T}}
\newcommand{\id}{\kw{ID}}

\newcommand{\nbr}[1]{\kw{$N_{#1}$}}
\newcommand{\nbre}[1]{\kw{$N_{#1}^E$}}
\newcommand{\adj}{\kw{Adj}}

\newcommand{\freq}{\mathcal{F}}
\newcommand{\cost}{\kw{Cost}}

\newcommand{\getnbrtype}{\kw{getOutNbr}}
\newcommand{\getnbretype}{\kw{getOutEdge}}
\newcommand{\glogedge}{\kw{getEdges}}
\newcommand{\glogvertex}{\kw{getFreq}}
\newcommand{\getcount}{\kw{getFreq}}
\newcommand{\getcandi}{\kw{getCands}}
\newcommand{\getcost}{\kw{getCost}}
\newcommand{\computecost}{\kw{computeCost}}

\newcommand{\gloguequery}{\kw{GLogueQuery}}

\newcommand{\planmap}{\kw{m}}
\newcommand{\lbound}{\kw{getLowerBound}}
\newcommand{\candi}{\kw{CandVTypes}}
\newcommand{\candie}{\kw{CandETypes}}

\newcommand{\init}{\kw{GreedyInitial}}

\newcommand{\code}[1]{\texttt{#1}}
\newcommand{\traits}{{\code{traits}}}
\newcommand{\trait}{{\code{trait}}}
\newcommand{\graphirbuilder}{\code{GIRBuilder}}
\newcommand{\physicalbuilder}{\code{PhysicalCostSpec}}
\newcommand{\physicalspec}{\code{PhysicalCostSpec}}
\newcommand{\converter}{\code{PhysicalConverter}}
\newcommand{\ir}{\kw{GIR}}
\newcommand{\logicalorder}{\kw{PatternOrder}}
\newcommand{\logicalorders}{\kw{PatternOrders}}

\newcommand{\wcoj}{\kw{WcoJoin}}

\newcommand{\kk}[1]{\texttt{#1}}
\newcommand{\scan}{{\kk{SCAN}}}                     
\newcommand{\expandedge}{{\kk{EXPAND}\_\kk{EDGE}}}  
\newcommand{\getvertex}{{\kk{GET}\_\kk{VERTEX}}}    
\newcommand{\expandpath}{{\kk{EXPAND}\_\kk{PATH}}}  
\newcommand{\pathstart}{{\kk{PATH}\_\kk{START}}}
\newcommand{\pathend}{{\kk{PATH}\_\kk{END}}}
\newcommand{\matchpattern}{{\kk{MATCH}\_\kk{PATTERN}}} 
\newcommand{\matchstart}{{\kk{MATCH}\_\kk{START}}}
\newcommand{\matchend}{{\kk{MATCH}\_\kk{END}}}
\newcommand{\project}{\kk{PROJECT}}                     
\newcommand{\select}{\kk{SELECT}}                       
\newcommand{\order}{\kk{ORDER}}                         
\newcommand{\limit}{\kk{LIMIT}}                         %
\newcommand{\optional}{\kk{OPTIONAL}}                   
\newcommand{\difference}{\kk{DIFFERENCE}}
\newcommand{\group}{\kk{GROUP}}                         
\newcommand{\joinopr}{\kk{JOIN}}                        
\newcommand{\union}{\kk{UNION}}                         %
\newcommand{\unfold}{\kk{UNFOLD}}                       

\newcommand{\map}{\kk{MAP}}
\newcommand{\flatmap}{\kk{FLATMAP}}
\newcommand{\source}{\kk{SOURCE}}
\newcommand{\expandvertex}{{\kk{EXPAND}}}
\newcommand{\expanddegree}{{\kk{EXPAND}\_\kk{DEGREE}}}

\newcommand{\topk}{\kk{TopK}} 

\newcommand{\expandinto}{\kw{ExpandInto}}
\newcommand{\expandintersect}{\kw{ExpandIntersect}}
\newcommand{\physicalexpand}{\kw{Expand}}
\newcommand{\hashjoin}{\kw{HashJoin}}
\newcommand{\physicalselect}{\kw{Select}}
\newcommand{\join}{\kw{Join}}
\newcommand{\expand}{\kw{Expand}}

\newcommand{\goptplan}{\kw{GOpt-plan}}
\newcommand{\gsplan}{\kw{GS-plan}}
\newcommand{\neoplan}{\kw{Neo4j-plan}}

\newcommand{\reffig}[1]{Fig.~\ref{fig:#1}}
\newcommand{\refsec}[1]{Section~\ref{sec:#1}}
\newcommand{\reftab}[1]{Table~\ref{tab:#1}}
\newcommand{\refalg}[1]{Algorithm~\ref{alg:#1}}
\newcommand{\refeq}[1]{Eq.~\ref{eq:#1}}
\newcommand{\refdef}[1]{Definiton~\ref{def:#1}}
\newcommand{\refthm}[1]{Theorem~\ref{thm:#1}}
\newcommand{\reflem}[1]{Lemma~\ref{lem:#1}}
\newcommand{\refex}[1]{Example~\ref{ex:#1}}
\newcommand{\refpro}[1]{Property~\ref{pro:#1}}
\newcommand{\refrem}[1]{Remark~\ref{rem:#1}}

\newcommand{\ie}{\emph{i.e.,}\xspace}
\newcommand{\eg}{\emph{e.g.,}\xspace}
\newcommand{\wrt}{\emph{w.r.t.}\xspace}
\newcommand{\aka}{\emph{a.k.a.}\xspace}
\newcommand{\kwlog}{\emph{w.l.o.g.}\xspace}
\newcommand{\kwhp}{\emph{w.h.p.}\xspace}

\newcommand{\todo}[1]{\textcolor{red}{$\Rightarrow$#1}}
\newcommand{\tbf}{\textbf{\textcolor{red}{X}}\xspace}
\newcommand{\warn}[1]{{\color{red}{#1}}}
\newcommand{\revise}[1]{{\textcolor{blue}{#1}}}
\newcommand{\lourevise}[1]{{\textcolor{orange}{#1}}}

\newcommand{\oom}{\texttt{OOM}\xspace}
\newcommand{\ot}{\texttt{OT}\xspace}
\newcommand{\rbo}{\kw{RBO}\xspace}
\newcommand{\cbo}{\kw{CBO}\xspace}

\newcommand{\hybrid}{\emph{Hybrid Semantics}\xspace}
\newcommand{\arbitrary}{\emph{Arbitrary Types}\xspace}

\newcommand{\vtype}[1]{\textit{#1}\xspace}
\newcommand{\company}{Alibaba}

\newcommand{\gsgopt}{\kw{GraphScope-GOpt}}
\newcommand{\neogopt}{\kw{Neo4j-GOpt}}

\newcommand{\fundamental}{\kw{Fundamental}}
\newcommand{\enhanced}{\kw{Enhanced}}

\newcommand{\sptrule}{\kw{SortProjectTrans}} 
\newcommand{\ajtrule}{\kw{AggJoinTrans}} 
\newcommand{\dfrule}{\kw{DegFusion}} 
\newcommand{\pkirule}{\kw{PKIndex}} 
\newcommand{\mtoirule}{\kw{MTOIndex}} 
\newcommand{\lprule}{\kw{LateProject}} 

\newcommand{\joinrule}{\kw{PatternJoin}}
\newcommand{\bjrule}{\kw{BinaryJoin}}
\newcommand{\verule}{\kw{VertexExpansion}}

\newcommand{\ftjrule}{\kw{FilterIntoJoin}}
\newcommand{\filterrule}{\kw{FilterIntoPattern}}
\newcommand{\fusionrule}{\kw{EVFusion}}
\newcommand{\trimrule}{\kw{FieldTrim}}
\newcommand{\commonrule}{\kw{ComSubPattern}}
\newcommand{\selectpush}{\kw{FilterPushDown}}
\newcommand{\joinelimrule}{\kw{JoinToPattern}}

\newcommand{\strategy}{\code{Strategy}}
\newcommand{\strategies}{\code{Strategies}}
\newcommand{\rulestrat}{\code{RuleBasedStrategy}}
\newcommand{\patstrat}{\code{PatternStrategy}}

\newcommand{\girplan}{\code{GIRPlan}}
\newcommand{\girop}{\code{GIROp}}
\newcommand{\execop}{\code{ExecOp}}
\newcommand{\execops}{\code{ExecOps}}

\newtheorem{remark}{Remark}[section]

\maketitle

\section{Introduction}
\label{sec:introduction}

Graph databases have become indispensable tools for managing interconnected data across domains such as social networks, fraud detection, and recommendation systems. At the heart of these systems lies the need to efficiently execute \textbf{Complex Graph Patterns (\cgps)}, which combine graph pattern matching with relational operations like projection, aggregation, and filtering. While existing graph databases like Neo4j~\cite{neo4j} and GraphScope~\cite{graphscope} provide foundational support for \cgps, their monolithic architectures impose critical limitations: (1) tight coupling with single query languages (e.g., Cypher~\cite{cypher} or Gremlin~\cite{gremlin}), hindering cross-language interoperability, and (2) fragmented optimization strategies that lack integration of state-of-the-art techniques like worst-case optimal joins or high-order statistics. These constraints impede performance and flexibility in real-world applications, where evolving query requirements and industrial-scale datasets demand modular, graph-native optimization.

In our SIGMOD 2025 paper, "A Modular Graph-Native Query Optimization Framework"~\cite{lyu2024gopt}, we introduced \gopt, a unified optimization framework designed to address these challenges. \gopt~ decouples query parsing, optimization, and execution through a graph intermediate representation (\ir), enabling support for multiple query languages (Cypher, Gremlin) and seamless integration with diverse backend engines (Neo4j, GraphScope). By combining heuristic rule-based optimizations (RBO), automatic type inference, and a cost-based optimizer (CBO) leveraging high-order statistics, \gopt achieves significant performance gains -- up to 48.6$\times$ speedup on Neo4j and 78.7$\times$ on GraphScope.

This technical report expands on the SIGMOD work by delving into \gopt’s architectural nuances and practical implementations. Key contributions include:
\begin{enumerate}
    \item \textbf{In-Depth Optimization Strategies}: We detail \gopt’s rule-based optimizations (e.g., \filterrule, \joinrule), cost-model-driven physical operator selection, and hybrid join strategies (binary joins, vertex expansion) tailored for backend engines.
    \item \textbf{Semantic Adaptations}: We formalize \gopt’s handling of diverse pattern-matching semantics (homomorphism, edge-distinct) across query languages, ensuring correctness while preserving optimization opportunities.
    \item \textbf{Physical Interface for Multiple Backends}: We demonstrate how \gopt’s \converter~ allows engines to covert their own executable operators (e.g., Neo4j’s \code{ExpandInto}, GraphScope’s \code{ExpandIntersect}), enabling backend-specific optimizations without compromising modularity.
    \item \textbf{Industrial Validation}: Through a granular analysis of \gopt’s optimization impact on the LDBC Social Network Benchmark (SNB)~\cite{ldbc_snb}, we dissect how and why specific plan transformations -- such as pattern order refinement, common pattern removal, and aggregation/filter pushdown -- enable significant performance gains. These case studies provide actionable insights for the community, offering a blueprint for optimizing complex graph queries.
\end{enumerate}

The report is structured as follows: \refsec{preliminaries} gives background knowledge of graph pattern matching. \refsec{ir} and \refsec{architecture} revisit \gopt’s \ir~ abstraction and core architecture, respectively. We detail the optimization strategies in \refsec{optimization_strategy}, and their applications to optimizing both simple and complex graph queries in \refsec{simple-queries} and \refsec{complex-queries}. \refsec{converter} details the design principles of \gopt's physical integration interface, formalizing two backend-specific methodologies for seamless integration with Neo4j and GraphScope, tailored to their Java-native and distributed execution models, respectively.

By bridging the gap between academic advancements and industrial demands, \gopt~ establishes a new paradigm for graph query optimization—one that is modular, extensible, and graph-native. This report serves as a comprehensive guide for practitioners seeking to leverage \gopt’s full potential in real-world graph analytics pipelines.

\vspace*{-1ex}
\section{Preliminaries}
\label{sec:preliminaries}

\begin{table}[t]
  \small
  \centering
  \caption{Frequently used notations.}
  \label{tab:notations}
  \begin{tabular}{ p{0.2\linewidth} | p{0.7\linewidth}  }
  \hline
    \bf{Notation} & \bf{Definiton} \\
    \hline
    $G(V_G, E_G)$ & A data graph with $V_G$ and $E_G$ \\
    \hline
    $P(V_P, E_P)$ & A pattern graph with $V_P$ and $E_P$ \\
    \hline
    $\footnotesize{\nbr{G}}(v)$, $\footnotesize{\nbre{G}}(v)$ & Out neighbors and out edges of $v$ in graph $G$ \\
    \hline
    $\elemtype_G(v)$, $\elemtype_G(e)$ & The type of vertex $v$ and edge $e$ in graph $G$ \\
    \hline
    $\type_P(v)$, $\type_P(e)$ & The type constraint of vertex $v$ and edge $e$ in pattern $P$, can be \btype, \utype~or \unionall\\
    \hline
    $\freq_{P,G}$ & The number of mappings of pattern $P$ in graph $G$ \\
    \hline
  \end{tabular}
\end{table}

\subsection{The Definitions}
\label{sec:patrel}
{Data graph $G = (V_G, E_G)$ in this report adheres to the definition of the property graph model \cite{angles17}.
$V_G$ and $E_G$ are the sets of vertices and edges, where $|V_G|$ and $|E_G|$ represent the number of vertices and edges. }
{Given $u, v \in V_G$, $(u, v) \in E_G$ is an edge directed from $u$ to $v$, and all the out neighbors and out edges of $u$ are denoted as $\nbr{G}(v)$ and $\nbre{G}(v)$, respectively.}
{Each vertex or edge in $G$ is associated with a type,}
denoted as $\elemtype_G(v)$ and $\elemtype_G(e)$ respectively.
Both vertices and edges can carry properties, which are key-value pairs.
Note that if no ambiguity arises, we omit $G$ from the subscript in the notations, and so as the follows.
{Considering two graphs \(G_1\) and \(G_2\), we assert that \(G_2\) is a subgraph of \(G_1\), symbolized as \(G_2 \subseteq G_1\), if and only if \(V_{G_2} \subseteq V_{G_1}\), and \(E_{G_2} \subseteq E_{G_1}\).}

A pattern $P = (V_P, E_P)$ is a small \textit{connected} graph.
Note that if $P$ is not connected, matching the pattern is equivalent to taking the Cartesian product of
the matches for its connected components, which is naturally the problem of \cgps~ in the following.
Each vertex and edge in the pattern graph is associated with a set of types as a type constraint, denoted as $\type_P(v)$ and $\type_P(e)$, respectively.
Besides, predicates can be specified to vertices and edges in the pattern graph as well.

Finding matches of $P$ in $G$ involves identifying all subgraphs $G'$ in $G$ {where} $P$ can be mapped to $G'$ via a homomorphism preserving edge relations and type constraints.
{The mapping function $h: V_P \rightarrow V_{G'}$ ensures that $\forall e=(u, v) \in E_P$, there is a corresponding edge $(h(u), h(v)) \in E_{G'}$}.
Additionally, the types of vertices and edges in $G'$ must align with the type constraints specified in $P$:
{(1) $\forall v \in V_P, \elemtype_{G'}(h(v)) \in \type_P(v)$, and (2) $\forall e=(u, v) \in E_P, \elemtype_{G'}((h(u), h(v))) \in \type_P(e)$.}
The number of mappings of $P$ in $G$ is called \textit{pattern frequency}, {denoted as $\freq_{P,G}$, or simply $\freq_P$ when $G$ is clear.}

{A complex graph pattern, termed as \cgp, extends \bgps~with further relational operations}.
{Optimizing \cgps~is challenging due to their hybrid semantics.
A straightforward way is to first identify matches of $P$ in $G$, and then subject the matched subgraphs to the remaining relational operators in \cgps~ for further analysis, such as projecting the properties of matched vertices and edges and selecting the matched subgraphs that satisfy certain conditions.}

In the realm of pattern matching, the \textit{pattern matching order} is pivotal, significantly influencing the overall efficiency and execution time of the process.
Consider the matching order for a pattern  $P$ as a sequence of vertices  $v^{(1)}, v^{(2)}, \ldots, v^{(|V_P|)}$, where each $ v^{(i)} \in V_P $ and $ i < j $ indicates that in the matching order, vertex $ v^{(i)} $ is mapped to a vertex in the data graph $ G $ before vertex $ v^{(j)} $. 
We can enumerate matching orders for pattern $P$ by permuting the vertices in $V_P$.
Additionally, there may be different matching strategies for each order. For example:
\begin{itemize}
    \item Start with the first vertex $v^{(1)}$, and then iteratively match the subsequent vertices $ v^{(2)}, \ldots, v^{(|V_P|)}$  to form the complete match.
    \item First match the vertices $v^{(1)}\ldots v^{(i)}$, then match the vertices $v^{(i)}\ldots v^{(|V_P|)}$, where $ i \in [1, |V_P|] $, and finally join the matching results of these two parts to generate the final matches.
    \item A hybrid approach that combines the above two methods.
\end{itemize}

Each matching order, along with its associated matching strategies, is referred to as a \logicalorder.
For a given pattern $P$, different \logicalorders~guarantee the same final results based on the equivalent pattern transformation rule \joinrule~(detailed in \refsec{pattern-strategies}), but involve different execution costs.
In \refsec{pattern-strategies}, we will show how \gopt~determine the best \logicalorder~for a given \bgp~using cost-based optimization techniques.

\subsection{Pattern Matching Semantics}
\label{sec:semantics}

It is important to note that in our framework, pattern matching is conducted under homomorphism semantics, which is useful due to its transformability.
However, different query languages may adopt different semantics for pattern matching. Generally, the semantics can be categorized as follows:
\begin{itemize}

  \item \textbf{Homomorphism Semantics}: Allows duplicates of vertices or edges in matching results. 

  \item \textbf{Edge-Distinct Semantics}: Allows duplicate vertices but not edges.

  \item \textbf{Vertex-Distinct Semantics}: Allows duplicate edges but not vertices.

  \item \textbf{Vertex-Edge-Distinct Semantics}: Disallows duplicates of both vertices and edges.

\end{itemize}

\begin{remark}
  \label{remark:semantics}
  Different query languages may adopt different semantics for pattern matching, which can lead to different matching results.
  Currently, Apache TinkerPop adopts homomorphism semantics in Gremlin~\cite{gremlin}, while Neo4j uses edge-distinct semantics in Cypher~\cite{cypher}.
  In \gopt, we primarily focus on homomorphism semantics for pattern matching as it is the most general.
  When integrating with different query languages, we apply the corresponding filters to ensure the matching results are consistent with the semantics of the respective query language.
\end{remark}

\begin{figure}[t]
  \centering
  \includegraphics[width=0.95\linewidth]{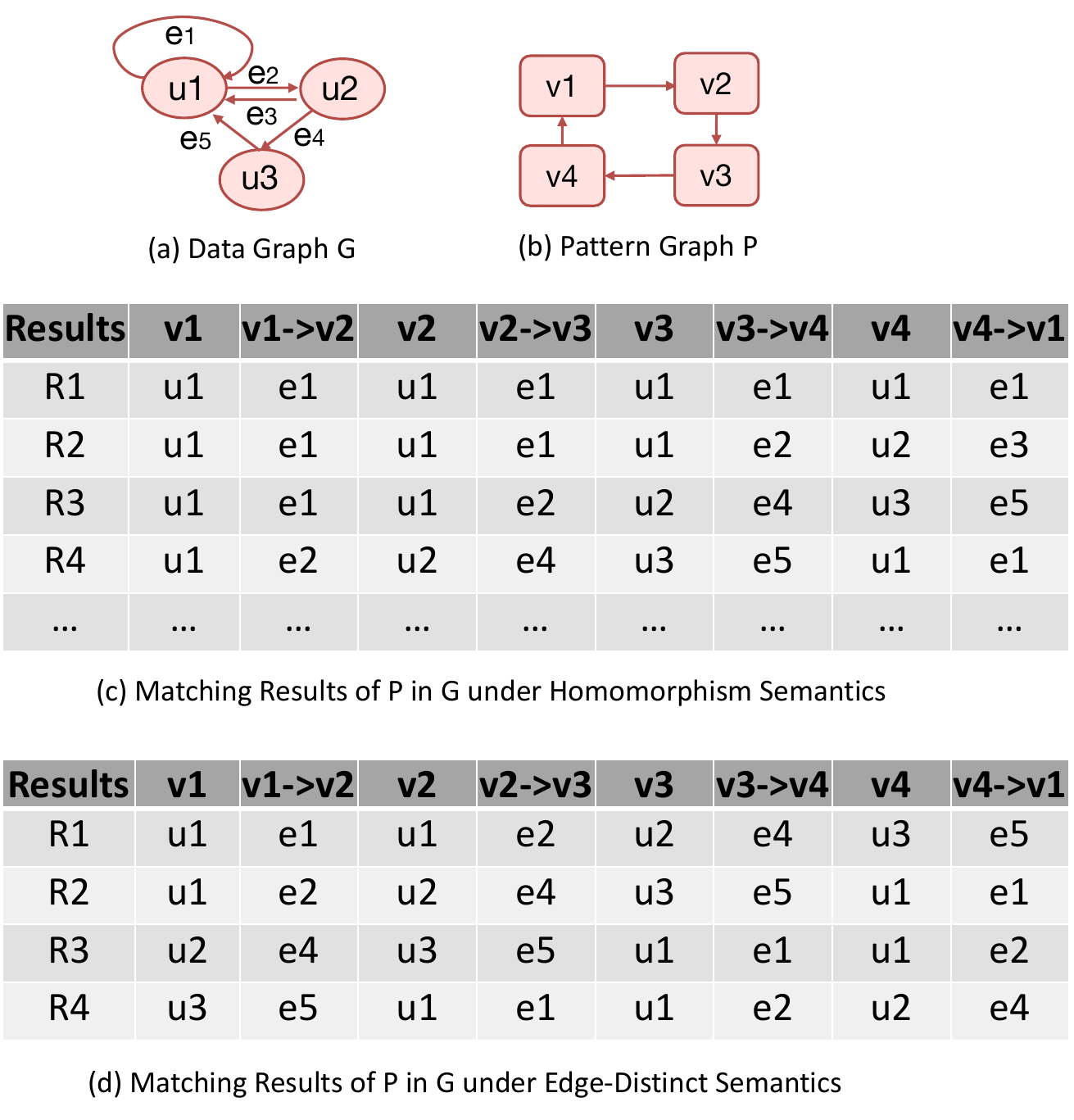}
  \caption{An example of pattern matching semantics.}
  \label{fig:semantics}
\end{figure}

\begin{example}

We illustrate the differences among various semantics in \reffig{semantics}, with a particular emphasis on homomorphism and edge-distinct semantics, which are employed by Gremlin and Cypher, respectively. 
\reffig{semantics}(a) presents a data graph $G$, while \reffig{semantics}(b) depicts a pattern $P$.
\reffig{semantics}(c) shows the pattern matching results of pattern $P$ within the data graph $G$ under homomorphism semantics. In this context, the matching results can include duplicated vertices and edges. For example, in $R_1$, all matched vertices are duplicates ($u_1$), and all matched edges are duplicates ($e_1$).
In contrast, under edge-distinct semantics, duplicate edges are not allowed. To comply with this, a further all-edge-distinct filter is applied to the results of \bgp~matching (i.e., the results in \reffig{semantics}(c)), to generate edge-distinct results in \reffig{semantics}(d).
\end{example}

\begin{figure}[t]
  \centering
  \includegraphics[width=0.95\linewidth]{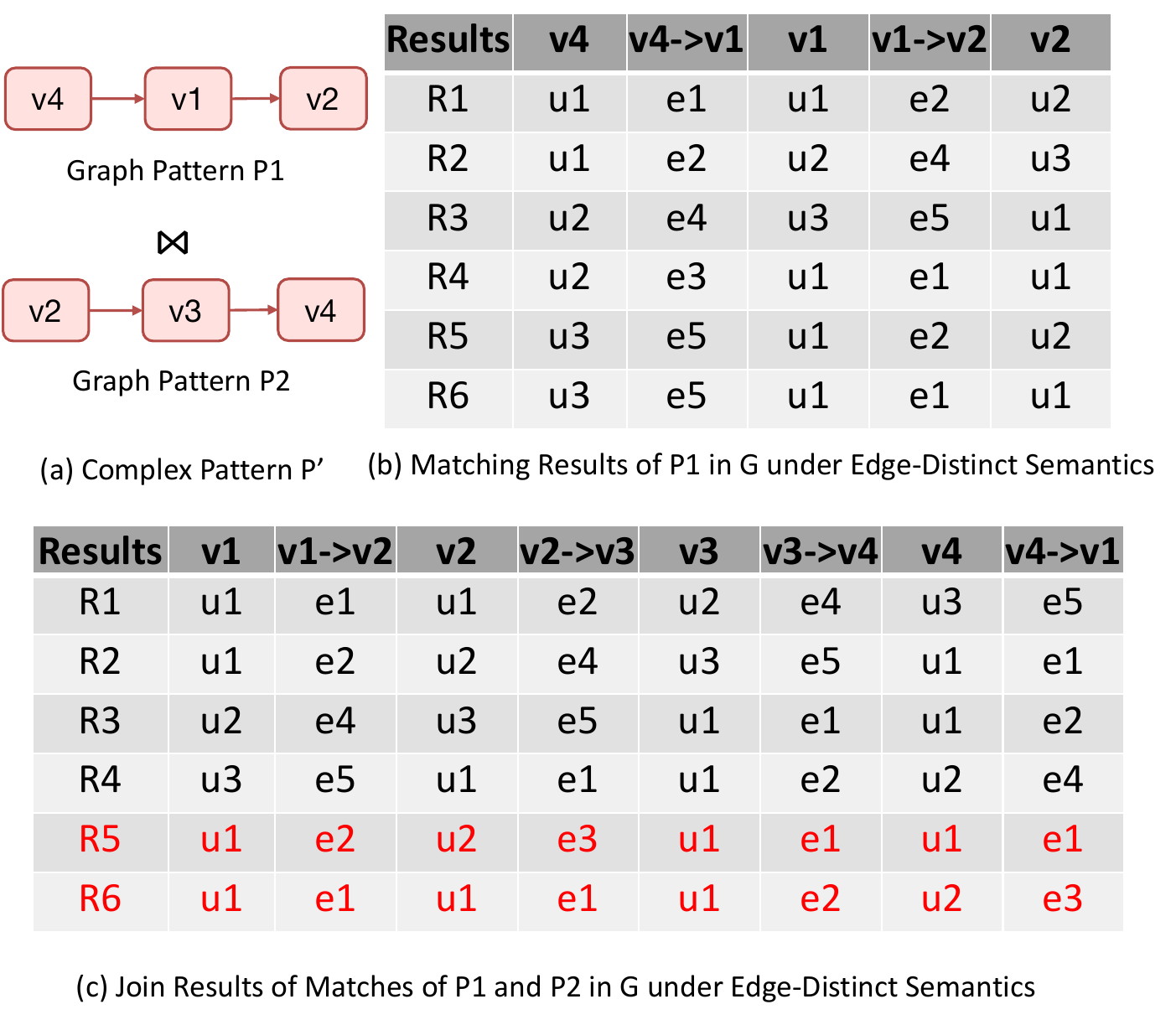}
  \caption{An example of joining two patterns under Edge-Distinct semantics.}
  \label{fig:joinelim}
\end{figure}

In our framework, we carefully design optimization rules to be consistent with the query language semantics. 
For example, the \joinelimrule~(which will be discussed in details in \refsec{joinelimrule}) aims to merge two \bgps~$P_1$ and $P_2$, initially connected by a \join~operator, into a single pattern $P$ where the join keys (i.e., vertices and/or edges) serves as the common vertices or edges in $P$.
Conversely, the \joinrule~(which will be discussed in details in \refsec{logical-orders}) attempts to decompose a pattern into two sub-patterns, match them separately, and then join the results to generate the final matches.
These two rules operate correctly under homomorphism semantics, making them suitable for Gremlin, since duplications are allowed in both pattern-based and join-based queries.
However, under non-homomorphism semantics, these rules must be applied with caution.
For example, when supporting Cypher, we cannot apply the \joinelimrule~to merge $P_1$ and $P_2$ into $P$, since the original join-based query results may contain duplicate edges, but after merging, the matchings of $P$ will not, which is inconsistent with the original results.
For the \joinrule~to decompose $P$ into $P_1$ and $P_2$, after matching $P_1$ and $P_2$ separately and joining the matching results, we will further apply an edge-distinct filter to the results to ensure no duplicate edges are present.

\begin{example}
  We illustrate an example of a join-based query under edge-distinct semantics in \reffig{joinelim}
  The \cgp~$P'$ joins two \bgps~$P_1$ and $P_2$, as shown in \reffig{joinelim}(a).
  The results of matching $P_1$ in $G$ (\reffig{semantics}(a)) are shown in \reffig{joinelim}(b). For brevity, the results for $P_2$ are omitted for brevity as they are the same to $P_1$'s.
  The results of \cgp~$P'$ in $G$ are displayed in \reffig{joinelim}(c), highlighting that joining the matching results of $P_1$'s and $P_2$'s can generate duplicate edges.
  
  First, let us consider the \joinelimrule. If we directly apply the \joinelimrule~to merge $P_1$ and $P_2$ into $P$ (\reffig{semantics}(b)), and match the merged pattern $P$ in $G$, we will eliminate duplicate edges in results as shown in \reffig{semantics}(d). This outcome is inconsistent with the original join-based query results shown in \reffig{joinelim}(c). Therefore, the \joinelimrule~should not be directly applied in Cypher.

  Next, let us consider the \joinrule. If we apply the \joinrule~to decompose $P$ (\reffig{semantics}(b)) into $P_1$ and $P_2$ (\reffig{joinelim}(a)), match them separately, and then join the results, we will obtain the results shown in \reffig{joinelim}(c). 
  However, the results contain duplicate edges, which contradict the matching results of $P$ in \reffig{semantics}(d) under edge-distinct semantics. 
  Therefore, we should apply an edge-distinct filter to the joined results (\reffig{joinelim}(c)) to ensure consistency with edge-distinct semantics.
\end{example}

\section{\ir~Abstraction}
\label{sec:ir}
In this section, we introduce the query-language-agnostic graph intermediate representation (\ir) for \gopt~to capture both graph and relational operations.
The \ir~ abstraction defines a data model $\mathcal{D}$ that describes the structure of the intermediate results during query execution, and a set of operators $\Omega$.

\subsection{Data Model}
The data model $\mathcal{D}$ presents a schema-like structure in which each data field has a String-typed name, accompanied by a designated datatype.
The supported datatypes encompass both graph-specific datatypes and general datatypes.
Graph-specific datatypes include \textit{Vertex}, \textit{Edge}, and \textit{Path}, as shown below:
\begin{itemize}
\item \textit{Vertex} is a datatype to represent the vertices in data graph.
It typically consists of:
\kw{ID} that serves as a unique identifier for the vertex;
\kw{TYPE} that characterizes the vertex class;
and \textit{properties} that includes property names and property values as a set of attributes associated with the vertex's type.
\item \textit{Edge} is a datatype to represent the edges in data graph.
It usually includes:
\kw{EID} that acts as a unique identifier for the edge, which is a triplet of \code{(edge\_id, src\_id, dst\_id)}, with \code{src\_id} and \code{dst\_id} pinpointing the source and destination vertices;
\kw{ETYPE} that represents the edge kind, which is also a triplet
 of \code{(edge\_type, src\_type, dst\_type)}, with \code{src\_type} and \code{dst\_type} specifying the source and destination vertex types;
and \textit{properties} that consist of property names and property values as a set of attributes associated with the edge's type.
\item \textit{Path} is a datatype of an array of vertices and edges that represents a sequence of connected vertices and edges in the data graph.
It is denoted as $p = [v_1, e_1, v_2, e_2, ..., v_n]$, where $v_i$ and $e_i$ are the $i$-th vertex and edge in the path respectively.
Specifically, \textit{Path} includes \kw{PID} as a unique identifier;
and a specific property of \code{length}, denoting the number of edges in the path.
\end{itemize}
General datatypes comprise \textit{Primitives} including \textit{Integer}, \textit{Float}, \textit{String} etc., and
\textit{Collections} representing a group of elements, e.g., \textit{List}, \textit{Set}, and \textit{Map}.
Notice that the properties in vertices and edges are of general datatypes.
For instance, a vertex with \textit{type} \code{PERSON} may have \textit{properties} of \code{name} (\textit{String}), \code{age} (\textit{Integer}), and \code{hobbies} (\textit{List}).

\subsection{Operators}
The operators in $\Omega$ operate on data tuples extracted from $\mathcal{D}$,
and produce a new set of data tuples as a result.
The set $\Omega$ is composed of graph operators and relational operators.
For clarity, we introduce only the parameters that are essential to understanding each operator's functionality, omitting those that are not critical to its core purpose.

\subsubsection{Graph Operators}
The graph operators are specifically for the retrieval of graph data. They include \getvertex, \expandedge, \expandpath, and \matchpattern.
\stitle{\getvertex}. The \getvertex~operator is designed to retrieve vertices from the data graph, or
to retrieve source or destination vertices from specified edges.

  \begin{description}
      \item[Parameters:] $(tag, alias, types, opt)$
      \begin{itemize}[leftmargin=0em, itemindent=-1.5em]
          \item $tag$: The identifier from which to obtain vertices. If it is tagged with an edge, it retrieves the source or target vertices from the tagged edge with the specified type constraints. If the $tag$ parameter is unspecified (NA), it retrieves vertices directly from the data graph.
          \item $alias$: A name to indicate the retrieved vertices in $\mathcal{D}$.
          \item $types$: The type constraints for the vertices.
          \item $opt$: The vertex option (if retrieving vertices directly
          from the data graph, this parameter will be ignored.):
          \begin{itemize}
            \item \emph{SRC}: Source vertices of the tagged edges.
            \item \emph{DST}: Destination vertices of the tagged edges.
            \item \emph{OTHER}: The other vertices of the tagged edges, which is in line with the following \texttt{\expandedge} operator when the edge is expanded in BOTH directions.
          \end{itemize}
      \end{itemize}
  \end{description}

\stitle{\expandedge}. The \expandedge~operator is used to retrieve edges from the data graph, or
to retrieve adjacent edges from specified vertices.
  \begin{description}
      \item[Parameters:] $(tag, alias, types, opt)$
      \begin{itemize}[leftmargin=0em, itemindent=-1.5em]
          \item $tag$: The identifier from which to expand edges. If it is tagged with vertices, it expands either outgoing (OUT) or incoming (IN) edges from the tagged vertices with specified type constraints. If the $tag$ parameter is unspecified (NA), it retrieves edges directly from the data graph.
          \item $alias$: A name to indicate the retrieved edges in $\mathcal{D}$.
          \item $types$: The type constraints for the edges.
          \item $opt$: The edge option (if retrieving edges directly
          from the data graph, this parameter will be ignored.):
          \begin{itemize}
            \item \emph{OUT}: Outgoing edges from the tagged vertices.
            \item \emph{IN}: Incoming edges to the tagged vertices.
            \item \emph{BOTH}: Both outgoing and incoming edges from the tagged vertices.
          \end{itemize}
      \end{itemize}
  \end{description}

  \stitle{\expandpath}. The \expandpath~operator is designed to retrieve paths from specified source vertices.
  \begin{description}
      \item[Parameters:] $(tag, alias, expand\_base, length, opt)$
      \begin{itemize}[leftmargin=0em, itemindent=-1.5em]
          \item $tag$: The identifier for the source vertices.
          \item $alias$: A name to indicate the retrieved paths in $\mathcal{D}$.
          \item $expand\_base$: A composite of \expandedge~and \getvertex~operators defining what is each hop in the path.
          \item $length$: The number of hops in the path expansion.
          \item $opt$: The path option:
          \begin{itemize}
              \item \emph{ARBITRARY}: No constraints.
              \item \emph{SIMPLE}: No repeated nodes in the path.
              \item \emph{TRAIL}: No repeated edges in the path.
          \end{itemize}
      \end{itemize}
  \end{description}

  \stitle{\matchpattern}. The \matchpattern~operator is used to describe a series of operations to match a complex pattern within the data graph.
  \begin{description}
      \item[Parameters:] $(expand\_base)$
      \begin{itemize}[leftmargin=0em, itemindent=-1.5em]
          \item $expand\_base$: A composite of \getvertex, \expandedge, and \expandpath~operators defining a pattern graph.
      \end{itemize}
  \end{description}

Notice that the $types$ parameter in the graph operators specifies type constraints to filter out desired classes of graph elements, where we allow either a single type constraint or a union of multiple type constraints, depending on the query requirements.
The $alias$ parameter indicates a name under which intermediate results are stored, allowing subsequent operations to reference these results using the $tag$ parameter.
We provide a special empty string tag to refer to the result of the immediate previous operation, thereby avoiding unnecessary data storage during execution.
Filter conditions can be integrated into these operators using optimization rules from \gopt, specifically the \filterrule.
In the \expandpath~operator, the number of hops is a specific positive integer. Ranged hop or even arbitrary hop (using Kleene star) will be handled in the future.
For the \matchpattern~operator, to facilitate illustration, we use \matchstart~and \matchend~to denote the beginning and end of a pattern match. Alternatively, within this paper, a \bgp~ may be depicted to represent the \matchpattern.

\subsubsection{Relational Operators}
We briefly discuss the essential relational operators in $\Omega$, which are widely used in RDBMS, including \project, \select, \order, \limit, \group, \unfold, \joinopr, and \union.

  \stitle{\project}. The \project~operator is used for projection, allowing the selection of specific columns and computation of additional values if necessary.
  \begin{description}
      \item[Parameters:] $(columns\_expressions)$
      \begin{itemize}[leftmargin=0em, itemindent=-1.5em]
          \item $columns\_expressions$: A list of columns or expressions to be included in the projection. The input columns (or those for these expressions) are specified by tags, and each output column can be given an alias for reference in subsequent operations.
      \end{itemize}
  \end{description}

  \stitle{\select}. The \select~operator filters rows (i.e., data) based on specific conditions in the data.
  \begin{description}
      \item[Parameters:] $(condition\_expression)$
      \begin{itemize}[leftmargin=0em, itemindent=-1.5em]
          \item $condition\_expression$: An expression defining the filter condition.
      \end{itemize}
  \end{description}

  \stitle{\order}. The \order~operator sorts the results according to specified columns and ordering options.
  \begin{description}
      \item[Parameters:] $(order\_by\_columns)$
      \begin{itemize}[leftmargin=0em, itemindent=-1.5em]
          \item $order\_by\_columns$: A list of columns, specified by tags, with options for sorting order (ASC or DESC).
      \end{itemize}
  \end{description}

  \stitle{\limit}. The \limit~operator restricts the number of results returned.
  \begin{description}
      \item[Parameters:] $(limit\_count)$
      \begin{itemize}[leftmargin=0em, itemindent=-1.5em]
          \item $limit\_count$: The maximum number of results to be returned.
      \end{itemize}
  \end{description}

  \stitle{\group}. The \group~operator groups the results by specific columns and applies aggregation functions.
  \begin{description}
      \item[Parameters:] $(group\_by\_columns, aggregation\_functions)$
      \begin{itemize}[leftmargin=0em, itemindent=-1.5em]
          \item $group\_by\_columns$: A list of columns, specified by tags, on which to group the results.
          \item $aggregation\_functions$: Functions applied for aggregation, such as COUNT, SUM, AVG, MIN, MAX, FIRST. Aggregated result can be assigned an alias for reference.
      \end{itemize}
  \end{description}

  \stitle{\unfold}. The \unfold~operator transforms a collection of nested results into a flat structure, where each element in the collection becomes a separate row.
  \begin{description}
      \item[Parameters:] $(collection\_column)$
      \begin{itemize}[leftmargin=0em, itemindent=-1.5em]
          \item $collection\_column$: The column, specified by a tag, containing the collection to be unfolded. The flattened results can be assigned an alias for reference.
      \end{itemize}
  \end{description}

  \stitle{\joinopr}. The \joinopr~operator is used to join the results of two sub-queries based on specific conditions, supporting multiple join types.
  \begin{description}
      \item[Parameters:] $(left, right, join\_keys, join\_type)$
      \begin{itemize}[leftmargin=0em, itemindent=-1.5em]
          \item $left$: The left sub-query for the join.
          \item $right$: The right sub-query for the join.
          \item $join\_keys$: The column keys, specified by tags, on which the join operation is performed, determining how rows from the two sub-queries are matched.
          \item $join\_type$: The type of join, such as INNER, LEFT OUTER, RIGHT OUTER, FULL OUTER, SEMI, or ANTI. The results can be assigned an alias for reference.
      \end{itemize}
  \end{description}

  \stitle{\union}. The \union~operator merges results from two sub-queries into a single result set.
  \begin{description}
      \item[Parameters:] $(left, right)$
      \begin{itemize}[leftmargin=0em, itemindent=-1.5em]
          \item $left$: The left sub-query to be merged.
          \item $right$: The right sub-query to be merged.
      \end{itemize}
  \end{description}

These operators can be applied on graph-specific data as well,
e.g., to project properties of vertices, to select edges with specific conditions,
to join two sub-paths into a longer one with the join key as the end vertices of the two sub-paths,
or to union the results of two sub-matching clauses.

\subsection{Applications of GIR}
We have designed \ir~ as a query-language-agnostic intermediate representation capable of expressing queries from diverse graph query languages.
Currently, we have implemented the construction of  \ir~\cite{graphscope_github} for two of the most popular graph query languages, Cypher~\cite{cypher} and Gremlin~\cite{gremlin}, enabling the broader application of optimization techniques based on \ir. As a result, with \gopt, GraphScope~\cite{graphscope} seamlessly supports queries from both query languages, optimizing them using the same unified set of optimization strategies that will be elaborated in the following sections.

\begin{figure}[t]
    \centering
    \includegraphics[width=1.0\linewidth]{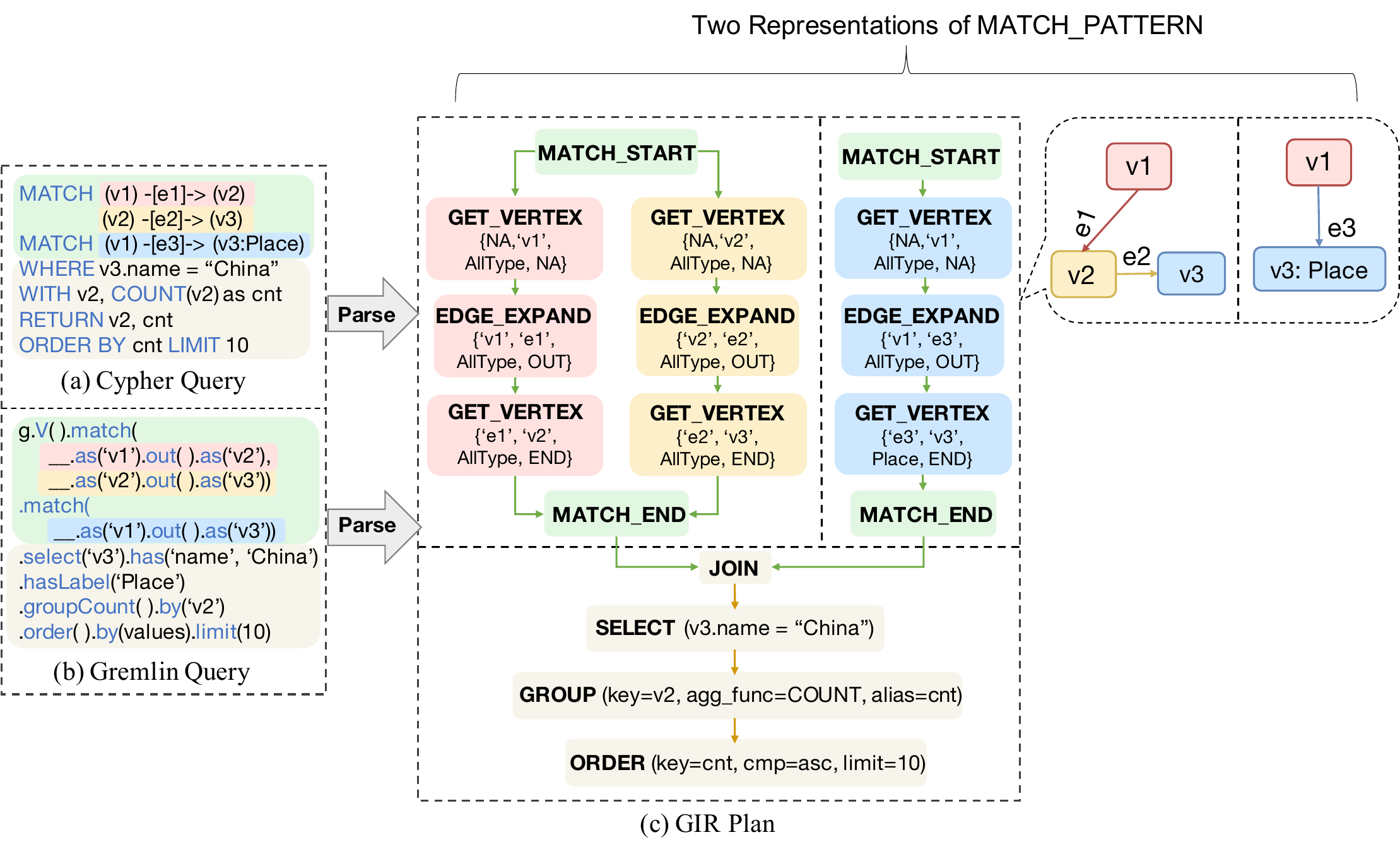}
    \caption{The \ir~Representations of Cypher queries. In the following, the \matchpattern~operator in an \ir~ will be illustrated as a graph view for simplicity.}
    \label{fig:gir}
\end{figure}

\reffig{gir} illustrates the \ir~ representation of the query example in Cypher and Gremlin languages. In \reffig{gir}, each \code{Match} clause is transformed into a \matchpattern~operator, which captures the vertex-edge relationships defined nested in the pattern. The two \code{Match} clauses are organized into a \join~operator (with an INNER join type), which combines the results of each \matchpattern. Notably, if the second \code{Match} clause is \code{OPTIONAL}, which means the pattern matching is optional,
the \join~operator will be the LEFT OUTER join type. The relational clauses following the \code{Match} clause are then converted into corresponding relational operators of \select, \group, and \order.

\begin{remark}
  {The definition of \ir~is inspired by existing work on graph relational algebra~\cite{Szrnyas2018ReducingPG}
  {and is enhanced with \matchpattern~to handle complex \bgps}.
  Moreover, \ir's development~follows an engineer-oriented approach, prioritizing support for commonly used functionalities over theoretical completeness. It {continuously} evolves to meet new requirements from real applications.}
\end{remark}

\section{Architecture}
\label{sec:architecture}

\begin{figure}[t]
    \centering
    \includegraphics[width=0.8\linewidth]{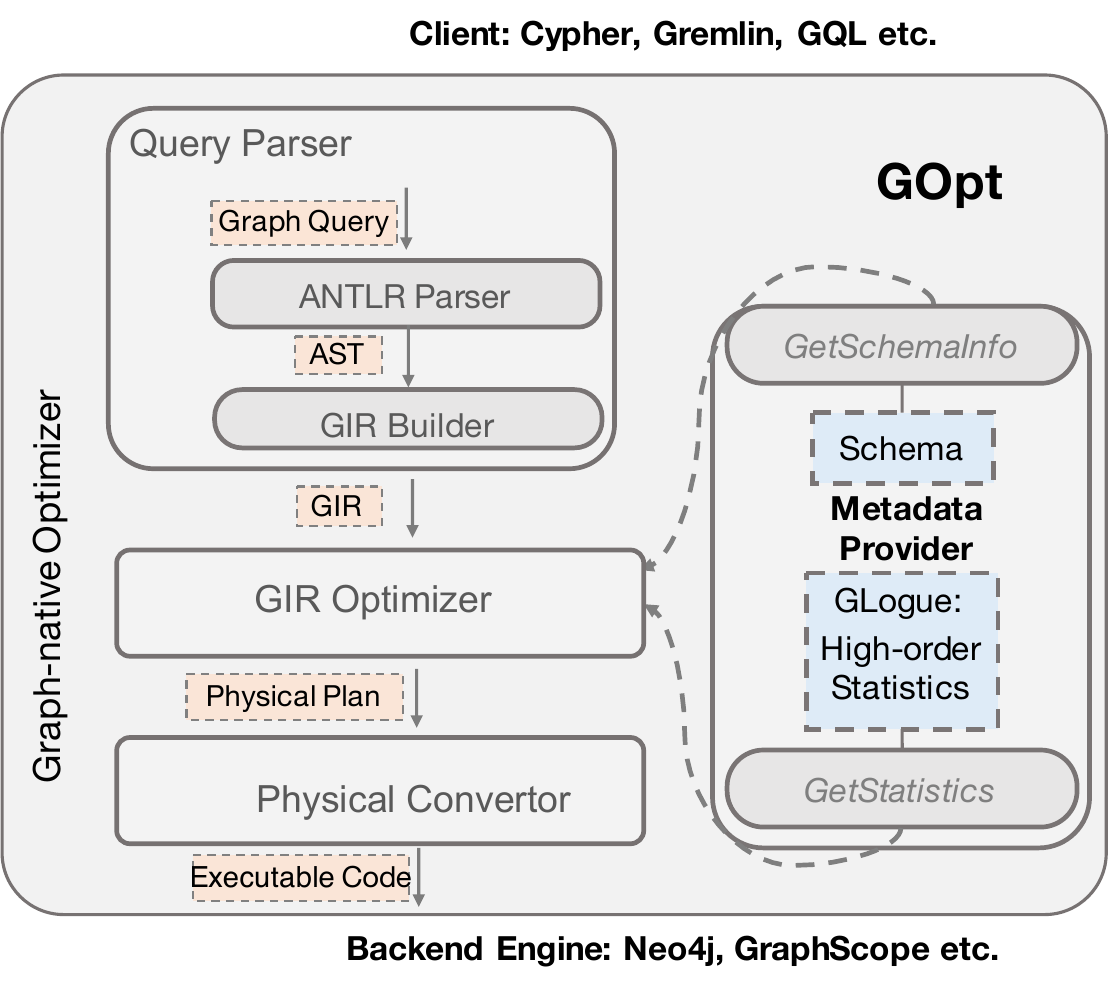}
    \caption{\gopt~ System Overview.}
    \label{fig:overview}
\end{figure}

The system overview of \gopt~ is illustrated in \reffig{overview}, which is composed of three primary layers: Query Parser, \ir~ Optimizer and Physical Converter. The Physical Converter provides a modular interface for integrating \gopt~ with different graph backends (detailed in Section~\refsec{converter}). In this section, we focus primarily on introducing the Query Parser and \ir~ Optimizer.

\subsection{Query Parser}
\label{sec:query_parser}

The Query Parser serves as the top layer of \gopt~to transform user queries into a logical plan in the \ir~format, to facilitate the following optimization process.

First, the parser performs syntax validation on the input graph queries using the ANTLR Tool~\cite{antlr}.
For queries that do not comply with the syntax rules, ANTLR errors are thrown immediately.
As aforementioned, \gopt~supports both Cypher and Gremlin.
The current implementation focuses on the key components of the query language, primarily involving query clauses for graph pattern matching and relational operations. The supported grammars can be found in~\cite{supported_grammars}.

Next, for valid queries, the \graphirbuilder~tool is used to construct the \ir~plan.
The \ir~plan represents a unified intermediate structure that is independent of the query syntax and serves as the foundation for all optimizations in \gopt.
We provide a code snippet to demonstrate how to construct the logical plan in \reffig{gir} using the \graphirbuilder:
\begin{lstlisting}
    GraphIrBuilder irBuilder = new GraphIrBuilder();
    pattern1 = irBuilder.patternStart()
      .getV(Alias("v1"),AllType())
      .expandE(Tag("v1"),Alias("e1"),AllType(),Dir.OUT)
      .getV(Tag("e1"),Alias("v2"),AllType(),Vertex.END)
      .expandE(Tag("v2"),Alias("e2"),AllType(),Dir.OUT)
      .getV(Tag("e2"),Alias("v3"),AllType(),Vertex.END)
      .patternEnd();
    pattern2 = irBuilder.patternStart()
      .getV(Alias("v1"),AllType())
      .expandE(Tag("v1"),Alias("e3"),AllType(),Dir.OUT)
      .getV(Tag("e3"),Alias("v3"),BasicType("Place"),Vertex.END)
      .patternEnd();
    query = irBuilder.join(pattern1,pattern2,
        Keys([Tag("v1"), Tag("v3")]), JoinType.INNER)
      .select(Expr("v3.name='China'"))
      .group(Keys(Tag("v2")),AggFunc.COUNT, Alias("cnt"))
      .order(Keys(Tag("cnt")), Order.ASC, Limit(10));
\end{lstlisting}

Here, \code{Alias()} defines an alias for results, accessible via \code{Tag()} in later operations. With the \graphirbuilder, the logical plan is constructed in a language-agnostic manner, facilitating the subsequent optimization process. Henceforth, when mention optimizing a query, we are referring to optimizing based on its \ir~.

\subsection{GIR Optimizer}
\label{sec:gir_optimizer}

\begin{figure}[t]
    \centering
    \includegraphics[width=1.0\linewidth]{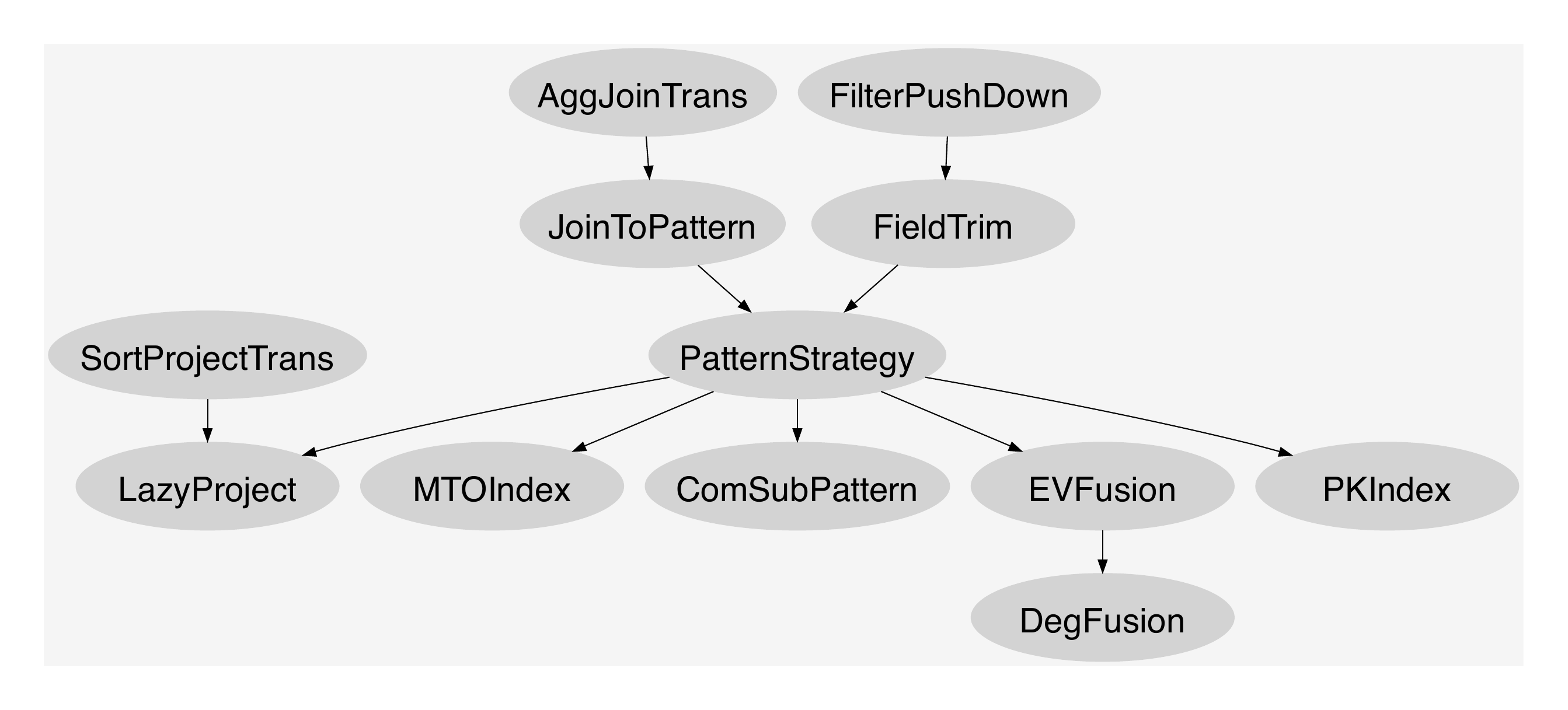}
    \caption{DAG of Optimization Process.}
    \label{fig:dag}
\end{figure}

In this phase, \gopt~ unfolds the \ir~structure, composed of a hybrid of patterns and relational operations, and transforms it into a physical plan that can be executed by the engine. The intermediate structure generated during the transformation is referred to as the \ir~plan. The input to this phase is the intermediate representation (\ir~) of the query, and the final output is the optimized \ir~ plan, which is ready for integration with the underlying engine.
The entire optimization process within \gopt~ can be conceptualized as a Directed Acyclic Graph (DAG), where each node embodies certain optimization \strategy. A \strategy~ encapsulates the application of either a single rule or a group of related rules within \gopt. The edges in the DAG represent dependencies between strategies, similar to the optimization phases in traditional database  systems, where certain rules depend on the outputs of others. \gopt~ applies the strategies in the topological order of the DAG, ensuring that the applications of strategies does not conflict with one another. The DAG formed by the various strategies that have been adopted by \gopt~ is shown in \reffig{dag}.

At the core of \gopt~, there lies the \strategy~ interface that guides the optimization process of each node in the DAG as follows:

\begin{lstlisting}
    interface Strategy {
        GIRPlan transform(GIRPlan input);
    }
\end{lstlisting}



The \code{transform} function defines how the input \ir~plan is transformed into the optimized \ir~plan. Depending on the implementation of \code{transform}, optimizations can be executed based on heuristic or top-down search approaches. Internally, \gopt~ provides two main types of \strategy~ implementations -- \rulestrat~ and \patstrat.

\rulestrat~ includes a series of heuristic rules, where each rule specifically implements the \code{transform} function to perform equivalent transformations on the input \ir~plan and output the transformed \ir~plan. These rules are partly reused from Calcite~\cite{Begoli_2018}, including: \ftjrule~\cite{ftjrule}, \trimrule~\cite{trimrule}, \sptrule~\cite{sptrule}, and \ajtrule~\cite{ajtrule}. Additionally, to handle specific optimizations related to graph data and operations, GOpt has implemented specialized rules for graph data models, including: \filterrule~, \joinelimrule~, \commonrule~, \fusionrule~, \dfrule~, \pkirule~, and \lprule. We will discuss these rules in detail in Section \ref{sec:strategies}.

\begin{figure}[t]
    \centering
    \includegraphics[width=0.8\linewidth]{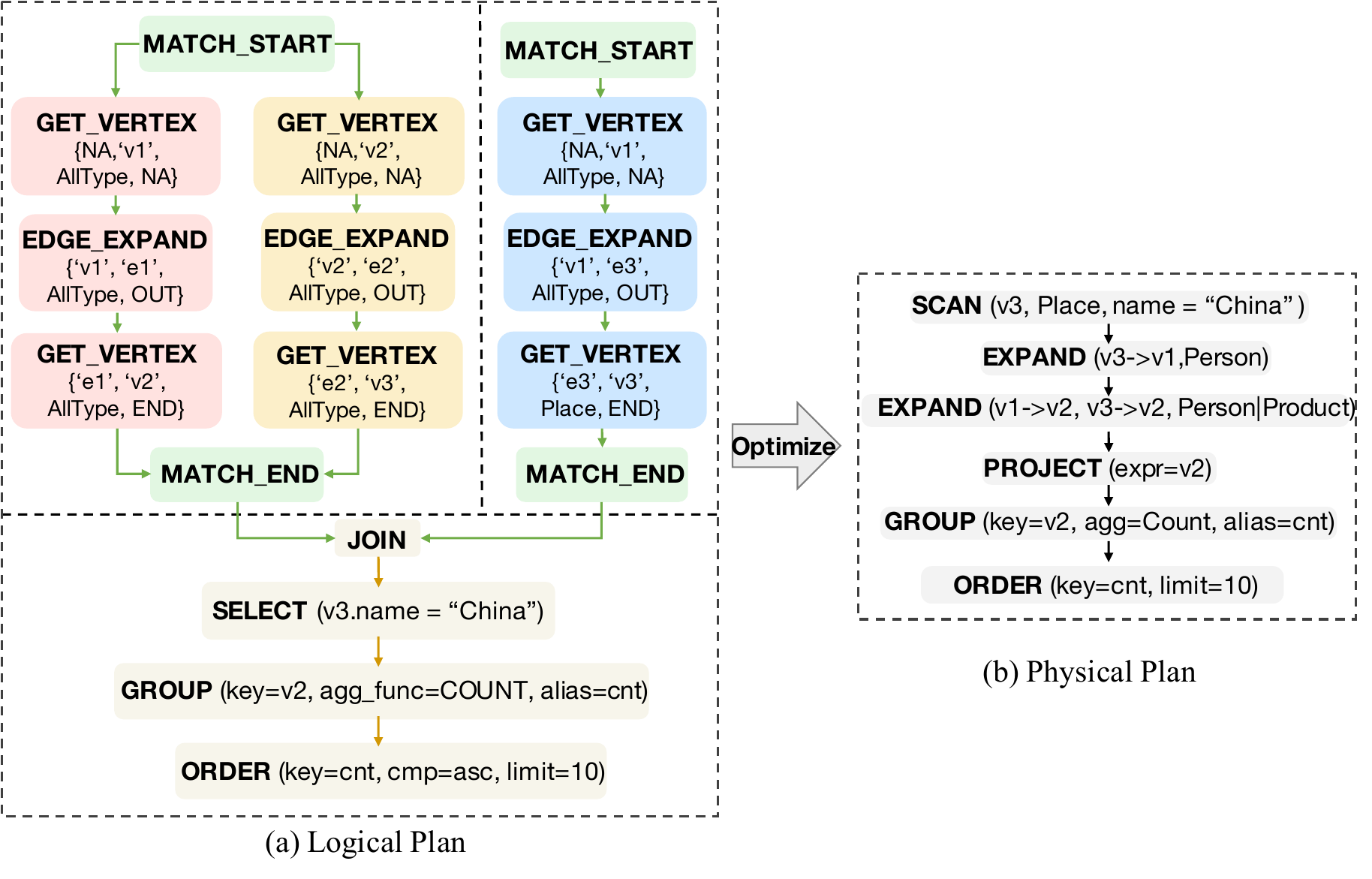}
    \caption{The GIR Optimization.}
    \label{fig:optimize}
\end{figure}

\patstrat~ primarily optimizes the execution sequence of graph operators within Patterns, referred to as \logicalorders~. The \code{transform} function takes a \ir~plan as input, composed of \matchpattern~ and other relational operators, as shown in \reffig{optimize}(a); it outputs an optimized \ir~plan representing the \logicalorder~, consisting of a series of physical operators, as shown in \reffig{optimize}(b). The \code{transform} function executes a top-down search algorithm as described in the \gopt~ paper~\cite{lyu2024gopt}, obtaining a series of \logicalorders~ along with their respective costs, and selects the \logicalorder~ with the lowest cost as the most optimized \ir~plan,

\section{Details of Optimization Strategies}
\label{sec:strategies}

In this section, we provide a detailed overview of the optimization strategies implemented in \gopt, which consists of two core components: \rulestrat~ and \patstrat.

The \rulestrat~ encompasses a series of heuristic rules. \gopt utilizes a heuristic-driven approach where these rules are applied immediately upon meeting specific preconditions, thus removing the necessity for cost-based estimation.
Some of these rules are adopted from Calcite, such as \ftjrule~\cite{ftjrule}, \trimrule~\cite{trimrule}, \sptrule~\cite{sptrule}, and \ajtrule~\cite{ajtrule}. Given that these rules have concrete implementations within Calcite, we will not delve into their introductions here.
Furthermore, to address particular optimizations pertinent to graph data and operations, \gopt has introduced specialized rules tailored for graph data models. These include \filterrule~, \joinelimrule~, \commonrule~, \fusionrule~, \dfrule~, \pkirule~, and \lprule. Detailed explanations of these rules will be provided in the following sections.


The second component, \patstrat, focuses on optimizing the order of graph operations within a single pattern. Unlike \rulestrat, the application of \patstrat~ relies on the availability of external metadata for cost estimation, which allows \gopt to determine whether to use a heuristic approach or a dynamic programming (DP) strategy for optimization. We will explore the implementation of \patstrat~ through three key interfaces: (1) \code{logicalOrder}: Determines the logical sequence of operations; (2) \code{physicalSpec}: Specifies the physical execution plan; (3) \code{getCost}: Estimates the cost of a given plan.

\subsection{\rulestrat}
\label{sec:common-strategies}

\subsubsection{\filterrule} 

\ftjrule~\cite{ftjrule} is a relational rule that implemented in Calcite, which aims to push filter conditions into a \join~ condition and into the inputs of the \join~. Built upon the idea, we implement \filterrule~ to further push down filter conditions into the graph operations nested in the Pattern. The following code snippet illustrates the preconditions before applying the rule:

\begin{lstlisting}
    public interface FilterIntoPatternConfig extends ... {
        FilterIntoPatternConfig DEFAULT =
          new Config()
              // Pattern is the input of Filter
              .withOperandSupplier(b0 ->
                  b0.operand(Filter.class).oneInput(b1 ->
                      b1.operand(AbstractLogicalMatch.class).anyInputs()));
    }
\end{lstlisting}

Considering the following example, we illustrate the transformation of this rule in detail. For the query:

\begin{lstlisting}
    Match (v1)-[e1]->(v2),
          (v2)-[e2]->(v3),
          (v1)-[e3]->(v3)
    Where v3.name = "China"
    Return v1.name, count(v2);
\end{lstlisting}

After the application of \filterrule~, the optimization effect is equivalent to rewriting the query as follows:

\begin{lstlisting}
    Match (v1)-[e1]->(v2),
    // the filter conditions has been pushed into the pattern
          (v2)-[e2]->(v3 {name: "China"}),
          (v1)-[e3]->(v3 {name: "China"})
    Return v1.name, count(v2);
\end{lstlisting}

\subsubsection{\joinelimrule} 
\label{sec:joinelimrule}
For two {\bgps}~connected by a \joinopr~operator, the \joinelimrule~aims to merge them into a single one, when the {join keys (i.e., vertices and/or edges)} serving as the common vertices and/or edges in the resulted {\bgp}.
This rule is effective under the homomorphism-based matching semantics as discussed in \refsec{semantics}.
It is applicable when the two \bgps~are connected by a \joinopr~operator with \code{JoinType.INNER}.
The following code snippet demonstrates the precondition checks:
\begin{lstlisting}
    public interface JoinToPatternRuleConfig extends ... {
        JoinToPatternRuleConfig DEFAULT =
          new Config()
          .withOperandSupplier(b0 ->
            b0.operand(Join.class)
              predicate(
                join -> join.getJoinType() == JoinType.INNER)
              )
              .inputs(
                b1 -> b1.operand(AbstractLogicalMatch.class).anyInputs(),
                b2 -> b2.operand(AbstractLogicalMatch.class).anyInputs());
    }
\end{lstlisting}

For example, consider the following query:
\begin{lstlisting}
    Match (v1)-[e1]->(v2),
          (v2)-[e2]->(v3)
    Match (v1)-[e3]->(v3)
    Return v1, v2, v3;
\end{lstlisting}
After applying the \joinelimrule~under homomorphism-based matching semantics, the two \bgps~are merged into a single one, resulting in the following query:
\begin{lstlisting}
    Match (v1)-[e1]->(v2),
          (v2)-[e2]->(v3),
          (v1)-[e3]->(v3)
    Return v1, v2, v3;
\end{lstlisting}

\subsubsection{\commonrule} 
For two {\bgps}~connected by binary operators such as \union~and \joinopr, we design the \commonrule~to identify common subpatterns in the two \bgps, and save the computation cost by matching the common subpattern only once.
This rule is applicable when the two \bgps~are connected by a binary operator, such as \union~or \joinopr, as shown in the following code snippet (\union~as an example, and \joinopr~is similar):
\begin{lstlisting}
    public interface ComSubPatternRuleConfig extends ... {
        ComSubPatternRuleConfig DEFAULT =
          new Config()
          .withOperandSupplier(b0 ->
            b0.operand(UNION.class))
              .inputs(
                b1 -> b1.operand(AbstractLogicalMatch.class).anyInputs(),
                b2 -> b2.operand(AbstractLogicalMatch.class).anyInputs());
    }
\end{lstlisting}
By identifying the \bgps~connected by a \union~or \joinopr, the \commonrule~will further detect the common subpattern in the two \bgps, and merge the common subpattern into a single one if it exists.
For example, consider the following query:
\begin{lstlisting}
    Match (v1:PERSON)-[]->(v2:PERSON)-[]->(v3:PLACE)
    Union
    Match (v1:PERSON)-[]->(v2:PERSON)-[]->(v4:COMMENT)
    Return v1, v2, v3, v4;
\end{lstlisting}
After applying the \commonrule, it identifies the common subpattern \code{(v1:PERSON)-[]->(v2:PERSON)} in the two \bgps, and merges the common subpattern into a single one and computes it only once, resulting in the following query:
\begin{lstlisting}
    Match (v1:PERSON)-[]->(v2:PERSON)
    with v1, v2
    Match (v2)-[]->(v3:PLACE)
    Union
    Match (v2)-[]->(v4:COMMENT)
    Return v1, v2, v3, v4;
\end{lstlisting}

\subsubsection{\fusionrule} 

The \fusionrule~is a specific optimization implemented for graph operators within patterns.
Traversal is one of the most commonly used query patterns in graph databases, typically composed of multi-hop expansions along different edge and vertex types. Each hop generally involves two main operations: (1) \expandedge~: Expanding edges of a specific type from the starting vertex, which may include edge filtering operations. (2) \getvertex~: Extracting the target vertex of a specified type from the edges, which may also involve further filtering of the target vertices.
However, in practical graph storage systems, target vertices are stored alongside their corresponding edges. This allows the operators of \expandedge~ and \getvertex~ to be merged into a new operator, \expandvertex~, which serves as the motivation behind our rule design.

Before applying this rule, several constraints must be satisfied:
(1) No alias operations on \expandedge~: Aliases typically indicate that subsequent operators will perform additional operations on the edges generated by \expandedge~ operator. If any alias remains after applying the \trimrule, the edges must be preserved separately and cannot be merged.
(2) No further type filtering on the target vertex: The target vertex’s type must be directly inferable from the pair types of the source vertex and the expand edge. For example, in \code{(a:PERSON)-[b:KNOWS]->(c:PERSON)}, according to the LDBC schema specification~\cite{ldbc_snb}, the type of vertex \code{c} can be determined as \code{PERSON} based on the pair types of \code{(a:PERSON)} and \code{(b:KNOWS)}, eliminating the need for additional type filtering. In contrast, consider \code{(a:PERSON)<-[:HASCREATOR]-(b:POST)}, \code{POST} and \code{COMMENT} are both valid target vertex types induced from the source vertex \code{(a:PERSON)} and the expand edge \code{[:HASCREATOR]}. In this case, the target vertex \code{b} requires further type filtering by \code{POST}, making the merging operation inapplicable.
(3) No additional property filtering on the target vertex: \expandvertex~ operator only supports filtering on edges, while separate property filtering on the target vertex would necessitate an extra filtering operation. Consequently, the merging operation would not yield any performance benefits in such cases.

\begin{lstlisting}
    public interface ExpandGetVFusionRuleConfig extends ... {
        ExpandGetVFusionRuleConfig DEFAULT =
          new Config()
          .withOperandSupplier(b0 ->
            b0.operand(ExpandE.class)
              .predicate(expandE -> noAlias(expandE))
              .oneInput(b1 -> b1.operand(GetV.class)
                              .predicate(getV -> noTypeFiltering(getV))
                              .predicate(getV -> noPredicateFiltering(getV))));
      }
\end{lstlisting}

In \reffig{degfusion_example}, we demonstrate how the \trimrule~and \dfrule~work together to optimize a query example.

\subsubsection{\dfrule} 

The \dfrule~ is an advanced fusion optimization that builds upon the \fusionrule~, specifically tailored for optimizing graph operators.
The \group~ operator, which functions as a reducer, can potentially become a performance bottleneck due to synchronization overhead. To address this, we have identified a specific pattern within \group~ operations where aggregations are performed over single edges. These operations can be transformed into the \expanddegree~ operator, effectively converting them into degree computations. This transformation allows the execution layer to utilize pre-maintained adjacent edge sizes, directly computing the results without incurring the performance penalties associated with the reducer operation.
The preconditions before applying the rule include:
(1) The GIR consists of an \expandvertex~ operator followed by an \group~ operator.
(2) The \group~ operation is semantically equivalent to computing the degree of the input \expandvertex~. Concretely, 
the condition checks whether the \group~ key is the starting vertex of the \expandvertex~ operator, and the \group~ function performs a distinct-count computation on the target vertices generated by \expandvertex~.

\begin{figure}[t]
    \centering
    \includegraphics[width=0.8\linewidth]{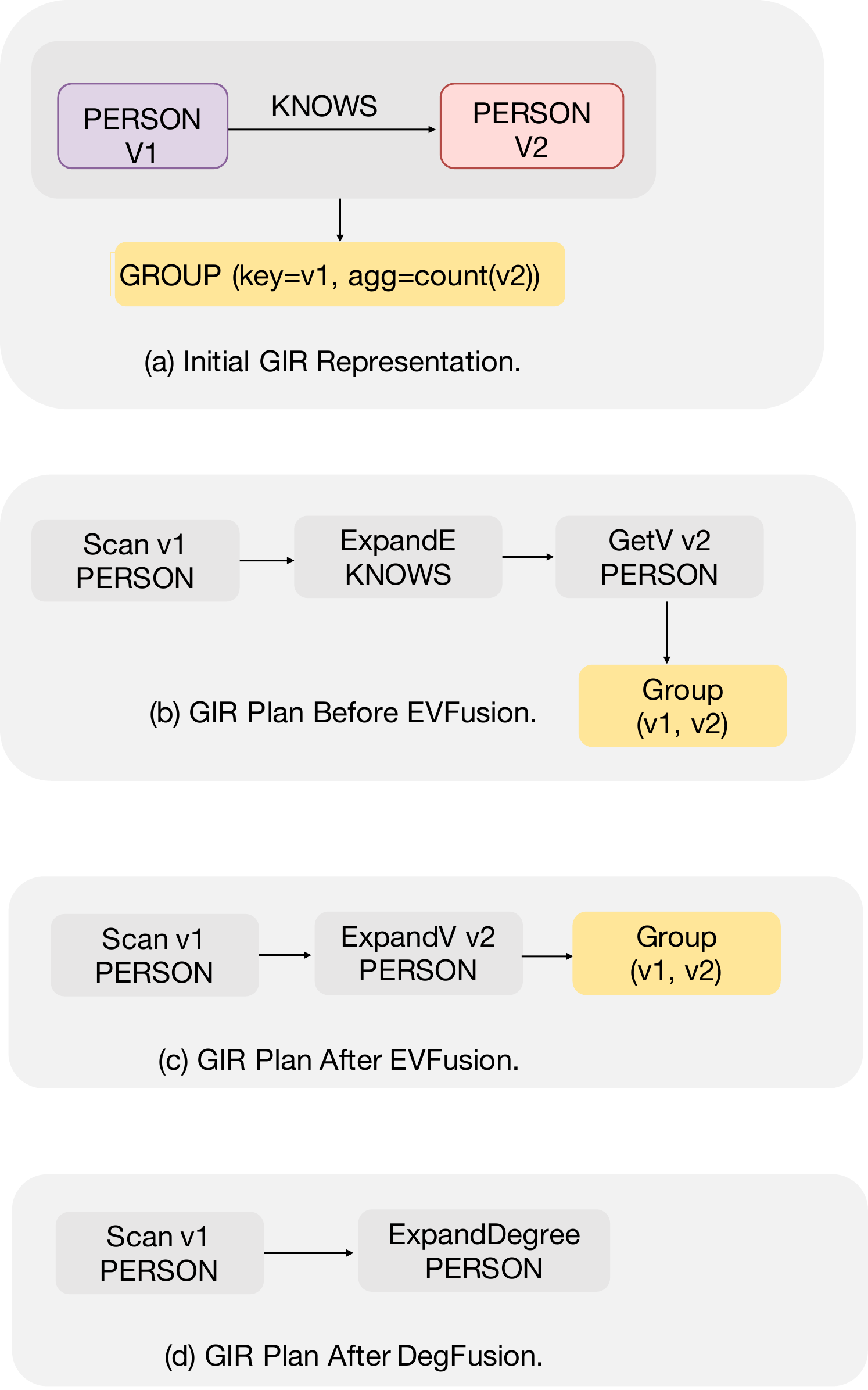}
    \caption{Transformations of \trimrule~ and \dfrule~.}
    \label{fig:degfusion_example}
\end{figure}

As shown in \reffig{degfusion_example}, we demonstrate the combined optimization effect of \trimrule~and \dfrule~on the following query.

\begin{lstlisting}
    Match (v1:PERSON)-[:KNOWS]->(v2)
    Return v1, count(distinct v2);
\end{lstlisting}

\subsubsection{\pkirule} 

In graph databases, it's common to maintain primary key indexes on vertices or edges to enhance the performance of graph-specific operations. We leverage the indexing capabilities provided by the execution layer to optimize graph operators through the use of primary key indexes. Given the following Cypher query:

\begin{lstlisting}
    Match (v1: PERSON) WHERE v1.id = 933 Return v1.name;
\end{lstlisting}

After application of the \filterrule~, the filtering condition will be fused into the source operation \code{(v1: PERSON)}, and is equivalent to rewriting the query as:

\begin{lstlisting}
    Match (v1: PERSON {id: 933}) Return v1.name;
\end{lstlisting}

The \code{(v1: PERSON {id: 933})} is represented by the physical operator \code{Scan v1(PERSON, id=933)} after \patstrat~.
Without a primary key index on the \code{id} attribute, the execution layer has to scan all vertices of type \code{PERSON} and then sequentially apply predicate filtering to identify the single vertex with \code{id=933}.
This approach can be inefficient, especially when dealing with large datasets. To optimize this, GOpt generates a hint by introducing the \code{IndexScan v1(PERSON, id=933)} operator, to guide the execution layer to use the primary key index on the \code{id} attribute, thereby optimizing the efficiency of \code{Scan} operation.


\subsubsection{\lprule} 

There are two primary strategies for retrieving property data in graph databases:
(1) Prefetch: In this approach, when a graph operator retrieves vertex or edge data, it simultaneously fetches the associated property data. During subsequent operations, if property data is needed, it can be accessed from cached values stored within the graph-specific data. This method reduces the latency of on-demand property data access and is particularly suitable for distributed scenarios. Prefetching helps mitigate costly communication overhead that would otherwise be incurred when accessing property data across machine nodes.
(2) Lazy Retrieval: This strategy defers the retrieval of property data until it is explicitly required by relational operations. When fetching graph-specific data types, the graph operator only retains minimal information, such as the internal ID or type for each vertex or edge. The key advantage of lazy retrieval is its ability to minimize the size of intermediate data during execution, which makes it ideal for single-node scenarios where there is no risk of additional communication overhead.

The \lprule~ is designed based on the concept of Lazy Retrieval. Currently, this rule is applied exclusively in single-node system settings. 

\subsection{\patstrat~}
\label{sec:pattern-strategies}
\subsubsection{Pattern Transformation}
\label{sec:logical-orders}
We optimize the query pattern by transforming it into various equivalent forms to facilitate the selection of the most efficient search order. The correctness of these pattern transformations is ensured by the \joinrule~as follows:

Given data graph $G$ and pattern $P_t$, with $P_{s_1}$ and $P_{s_2}$ where $P_t = P_{s_1} \bowtie_k P_{s_2}$ and $\bowtie_{k}$ is the join operator with join key $k = V_{P_{s_1}} \cap V_{P_{s_2}}$.
Let $R(P,G)$, or $R(P)$ for brevity, represent the {results of matching } $P$ in $G$.
{Under homomorphism-based matching semantics}, {$R(P_t)$} can be computed by:
\begin{equation}
R(P_t) = R(P_{s_1}) \bowtie_{k} R(P_{s_2})
\end{equation}

It should be noted that, for other non-homomorphic semantics, additional filters can be applied to the results of $R(P_t)$ to ensure compliance with the query semantics, as we have discussed in \refsec{semantics}.

Based on this equivalent rule in pattern matching, we introduce two build-in strategies in \gopt, to perform pattern transformations while ensuring the correctness:

\begin{itemize}[noitemsep,topsep=0pt]

    \item {\bjrule:} denoted as $\joinopr({P_{s_1}, P_{s_2}} \rightarrow P_t)$, this strategy decomposes a query pattern $P_t$ into two sub-patterns $P_{s_1}$ and $P_{s_2}$ where $P_t = P_{s_1} \bowtie_k P_{s_2}$. In this case, we can find matchings of $P_{s_1}$ and $P_{s_2}$ separately, and then using a binary join operator to join the results of $P_{s_1}$ and $P_{s_2}$ to generate the final results.

    \item {\verule:} denoted as $\expandvertex(P_s, P_v \rightarrow P_t)$, this strategy applies when $V_{P_v} = V_{P_t} \setminus V_{P_{s}} = \{v\}$. We can firstly find matchings of $P_s$, then match vertex $v$ by expanding edges directly from $P_s$'s matchings, to obtain the final results of $P_t$.

\end{itemize}

\begin{figure}[t]
    \centering
    \includegraphics[width=\linewidth]{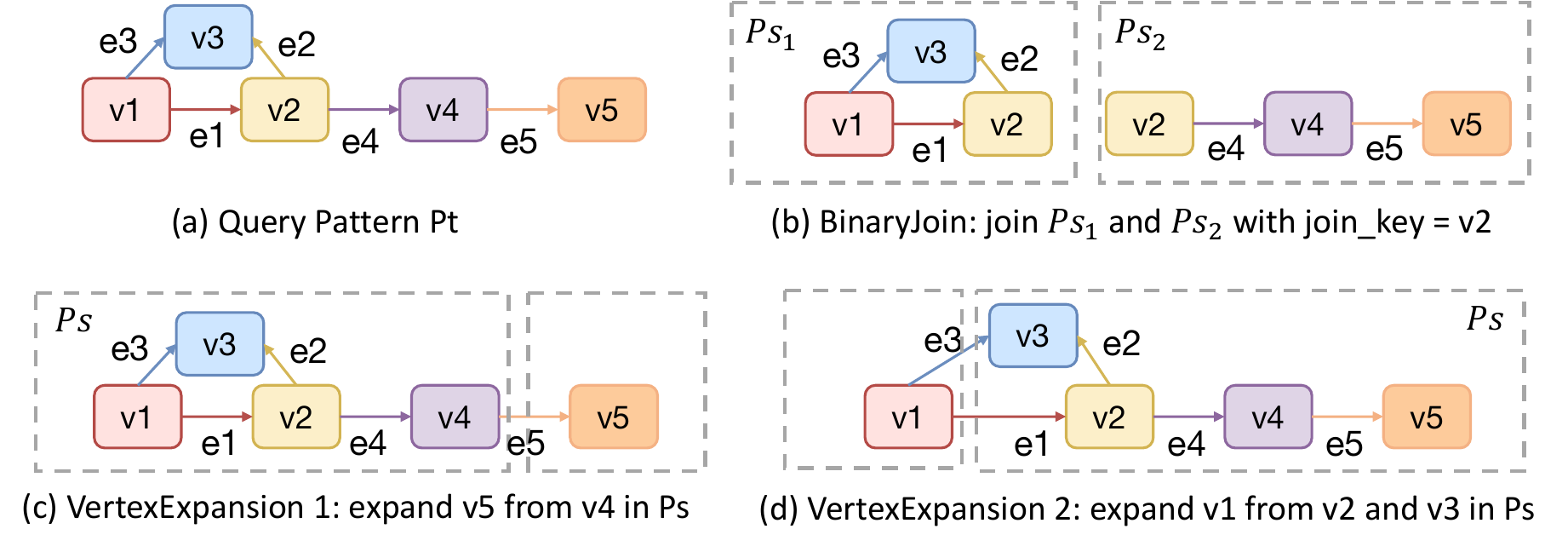}
    \caption{Examples of Pattern Transformation.}
    \label{fig:pattern_trans}
\end{figure}

\begin{example}
We show an example of the pattern transformation process in \reffig{pattern_trans}. The initial query pattern, $P_t$, is depicted in \reffig{pattern_trans}(a), while three equivalent transformations of $P_t$ are illustrated in \reffig{pattern_trans}(b)-(d).
In \reffig{pattern_trans}(b), the pattern results from applying the \bjrule~to decompose $P_t$ into two sub-patterns, $P_{s_1}$ and $P_{s_2}$, where $P_t = P_{s_1} \bowtie_k P_{s_2}$ and the join key $k = v_2$.
Figures \reffig{pattern_trans}(c) and \reffig{pattern_trans}(d) demonstrate the application of the \verule, expanding a single vertex from $P_s$ to $P_t$.
Specifically, in \reffig{pattern_trans}(c), a single edge, $e_5$, along with its adjacent vertex $v_5$, is expanded from $P_s$ to $P_t$,
while \reffig{pattern_trans}(d) illustrates that multiple edges, $e_1$ and $e_3$, with the common adjacent vertex $v_3$, are expanded from $P_s$ to $P_t$.
\end{example}

\subsubsection{\physicalbuilder}
Note that certain operations can have rather different computational costs in different engines. To address the issue, we introduce the \physicalbuilder~ interface to allow the backend engine to register their implementation costs for the pattern transformations.
Below, we demonstrate the details of \physicalbuilder~of the two different pattern transformation strategies, \bjrule~and \verule, when integrated with \gs~and Neo4j.

\begin{itemize}[noitemsep,topsep=0pt]
    \item \textcolor{black}{\kw{BinaryJoinCostSpec:}} The \joinopr~may have different implementations such as \hashjoin, \kw{NestedLoopJoin} and \kw{SortMergeJoin}. When integrated with \gs~and Neo4j, they both adopt the \hashjoin~implementation. So the cost of executing the \joinopr~is defined as:
    \begin{equation}
        \cost_{\join} = \alpha_{\join} \times (\freq_{P_{s_1}} + \freq_{P_{s_2}})
    \end{equation}

    where $\alpha_{\join}$ is a normalized factor for the \joinopr~operation, and so as the other operations.

    \item \textcolor{black}{\kw{VertexExpansionCostSpec:}}
    If the vertex $v$ only has one adjacent edge to $P_s$, i.e., $E(P_t)\setminus E(P_s) = \{e\}$, both Neo4j and \gs~implement it by getting the neighbors through the edge $e$.
    However, if the vertex $v$ has multiple adjacent edges to $P_s$, i.e., $E(P_t)\setminus E(P_s) = \{e_1, \ldots, e_n\}$, where $e_i = (v_i, v)$ and $v_i \in V(P_s)$, Neo4j and \gs~adopt different implementations.
    Specifically, Neo4j implements an \expandinto~operation,
    which first expands the edge $e_1$ from $v_1$ to get candidate set of matchings for $v$, and then expands the edge $e_2$ from $v_2$, filtering out those are not in the candidate set, and so on so forth until the last edge $e_n$, and finally returns the matchings of $P_t$. Therefore, the cost of executing the \expandvertex~operation when integrating Neo4j is defined as:
    \begin{equation}
        \label{eq:neo4j_expand_cost}
        \cost_{\expandvertex} = \alpha_{\expandvertex} \times \sum_{i=1}^{n} \freq_{P_i}
    \end{equation}
    In contrast, \gs~adopts a more efficient implementation, called \expandintersect, which is based on the worst-case optimal join algorithm. In this case, \gs~expands edges from $v_1, \ldots, v_n$ simultaneously, and then intersect the candidate sets to get the final matchings. The cost of executing the \expandvertex~operation when integrating \gs~is defined as:
    \begin{equation}
        \label{eq:gs_expand_cost}
        \cost_{\expandvertex} = \alpha_{\expandvertex} \times n \times \freq_{P_s}
    \end{equation}
\end{itemize}

With these cost estimations, along with those of other physical operators omitted here for brevity, \gopt~can employ a general cost model to select the most efficient physical plan by minimizing the total operator costs within the plan.
Please note that during pattern optimization, it is not necessary to specify the implementations of the operators in different backends, e.g., \expandinto~or \expandintersect, which, we refer them as \execops.
Instead, our focus is on the cost of executing these operators, which influences the physical plan \gopt~selects.
Once optimization is complete, \gopt~converts the optimized physical plan into a format executable by the integrated backend engine, which will then convert each \girop~to the corresponding \execop~of the backend engine. We will discuss this process in detail in \refsec{converter}.
Additionally, defining these cost functions is optional; if a cost function is not specified, \gopt~can still apply \rulestrat~to optimize the patterns.

\begin{figure}[t]
    \centering
    \includegraphics[width=0.9\linewidth]{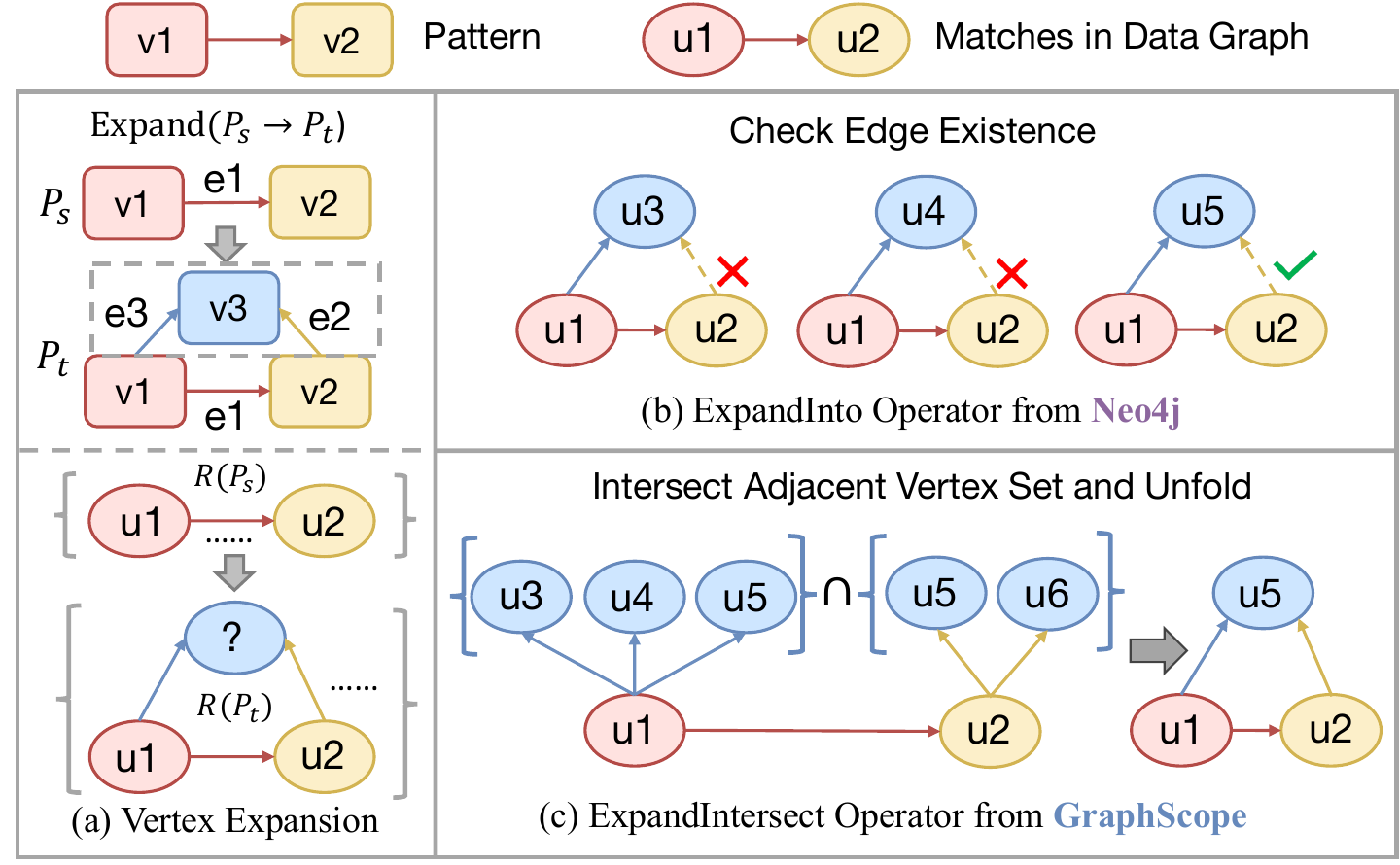}
    \caption{An Example of Vertex Expansion Implementations}
    \label{fig:cbo}
\end{figure}

\example{
    We illustrate the different vertex expansion implementations of Neo4j and \gs~ when multiple edges are expanded in \reffig{cbo}.
    Initially, pattern $P_s$ matches to $(u_1, u_2)$ in the data graph (\reffig{cbo}(a)). To match $P_t$ by expanding $e_3$ and $e_2$,
    Neo4j first expands $e_3$, yielding three mappings (\reffig{cbo}(b)), and then performs \expandinto~to find matches for $e_2$ connecting $u_2$ to $u_3$, $u_4$ and $u_5$, respectively. As \expandinto~flattens the intermediate matching results, the cost is the sum of frequencies of intermediate patterns, as shown in \refeq{neo4j_expand_cost}.
    In contrast, \gs~uses \expandintersect, which begins by finding the match set $R(P_1)$ by expanding $e_3$, yielding one intermediate result. It then expands $e_2$ and intersects the matched set with $R(P_1)$ to obtain $R(P_t)$, and finally unfolds the match set after expanding all the edges (\reffig{cbo}(c)). \expandintersect~reduces computation by avoiding flattening intermediate results, with the cost defined in \refeq{gs_expand_cost}.
}


\section{Optimizing Complex Queries}
\label{sec:complex-queries}

In this section, we demonstrate how query optimization can be achieved through the combination of multiple \strategies. We select several queries from IC and BI query sets to illustrate the optimization process.
For each query, we first register the \strategies~DAG with \gopt~. The DAG specifies the application order of the \strategies, which \gopt~ follows sequentially. Before applying a \strategy, \gopt~ evaluates whether the query satisfies the preconditions of the \strategy. If the preconditions are not met, the \strategy~is skipped, and the next \strategy~is applied. If the preconditions are satisfied, the \strategy~is applied to optimize the current \ir~ Plan, and its output serves as the input for the subsequent \strategy. The following examples provide a detailed view of how these queries are optimized.

\subsection{Single-Pattern Heuristic Optimization.}

\begin{figure*}[ht]
  \centering
  \includegraphics[scale=0.25]{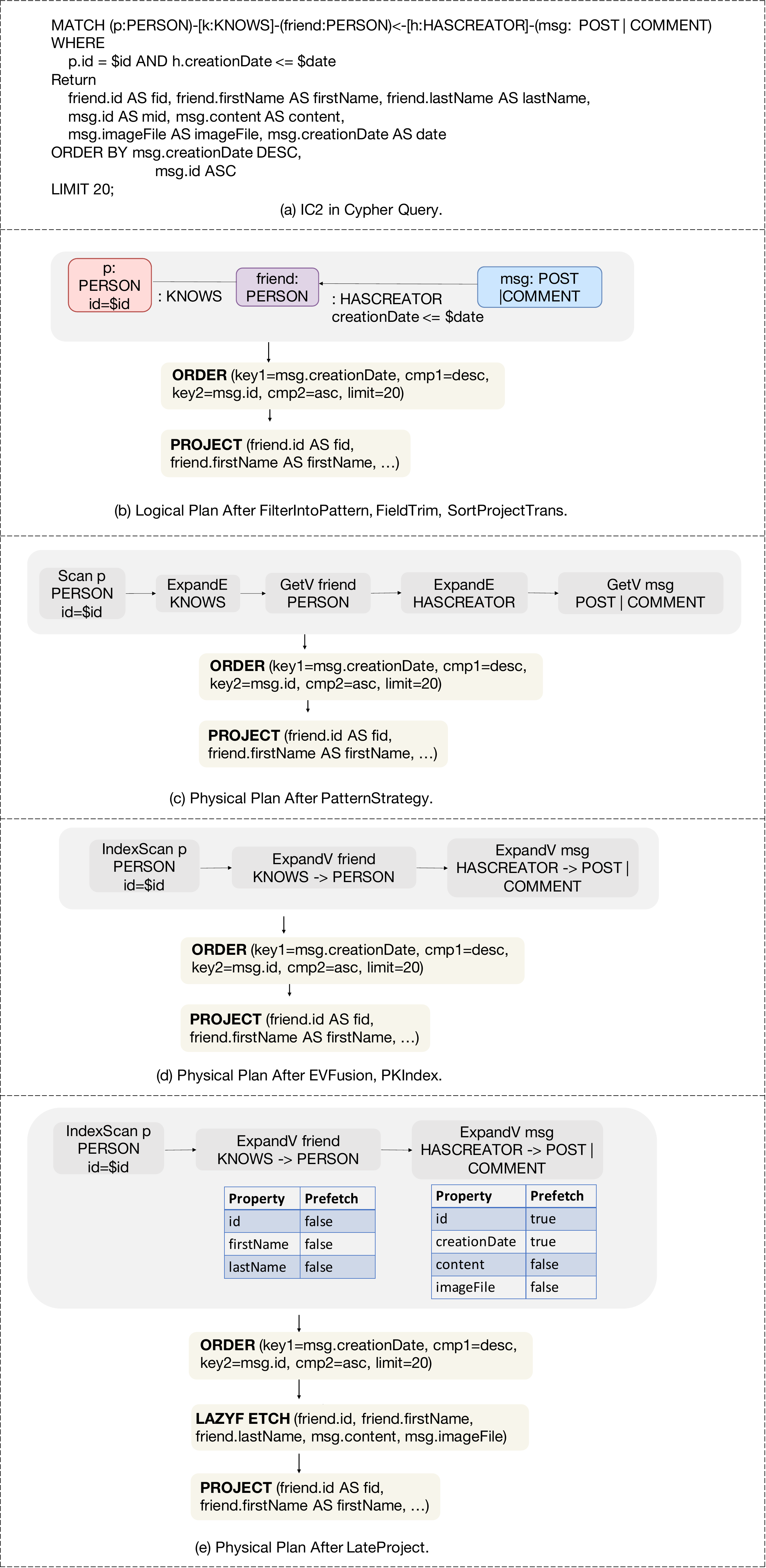}
  \caption{Optimization Process of $IC_2$.}
  \label{fig:rbo-ic2}
\end{figure*}

As illustrated in Figure\reffig{rbo-ic2}(a), the $IC_2$ query consists of a single MATCH clause combined with a series of relational clauses.
The query pattern represents a straightforward 2-hop traversal, where the \code{(p:PERSON)} node is filtered using a primary key lookup based on its property \code{id}.
The pattern itself does not involve complex optimizations, and the final \logicalorders~can be directly derived from the order specified in the user-provided query. The primary optimization focus lies in the combined application of \patstrat~ and other \rulestrat. During the optimization process, only a subset of the registered \strategies~is effectively applied, which are highlighted in blue in \reffig{ic2-dag}. \gopt~ follows the specified order of these \strategies~to apply the optimizations as follows:

\begin{figure}[t]
  \centering
  \includegraphics[width=1.0\linewidth]{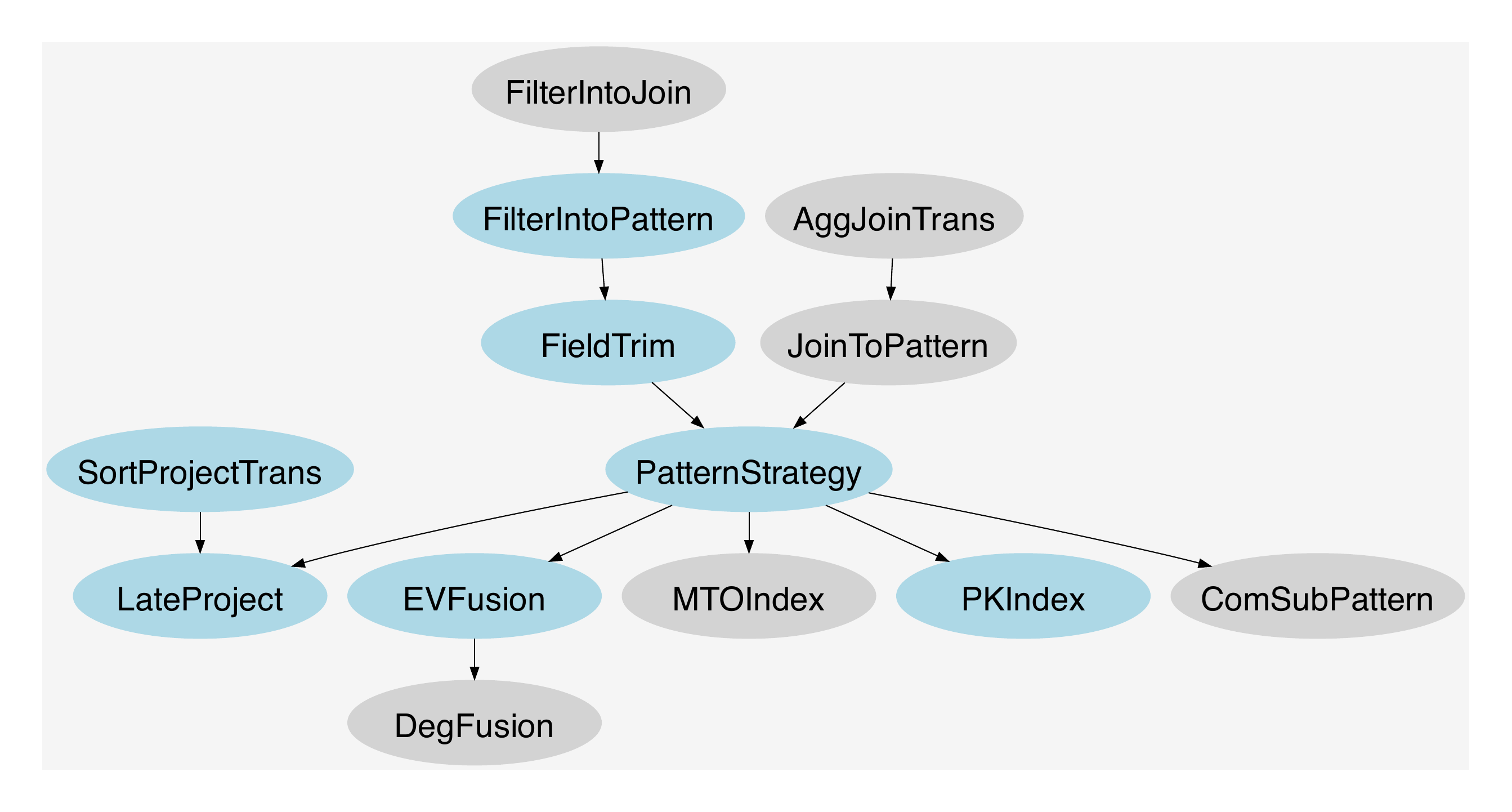}
  \caption{Strategies DAG of $IC_2$.}
  \label{fig:ic2-dag}
\end{figure}

\begin{itemize}
  \item \gopt~ translates the Cypher query into a \ir~ using the \graphirbuilder~tool. The \ir~ consists of a Pattern structure along with relational operations that correspond to the WHERE, RETURN, and ORDER clauses in the original Cypher query.
  Due to page size constraints, the initial structure of the \ir~ is omitted from the figure. The \ir~ after the application of \filterrule~, \trimrule~, and \sptrule~ is shown in \reffig{rbo-ic2}(b). By applying \filterrule~, the filtering conditions in the WHERE clause are pushed down into the MATCH clause, specifically onto the \code{(:PERSON)} node and the \code{(:HASCREATOR)} relationship.
  \trimrule~eliminates redundant columns generated by the MATCH clause, such as columns k and h derived from the \code{[:KNOWS]} and \code{[:HASCREATOR]} relationships, respectively. Additionally, this rule retains only the minimal required set of node and edge properties, which, when combined with \lprule~, maximizes optimization benefits, as will be discussed in the next step. By leveraging Calcite’s \sptrule~, the \topk~ operation—comprising \order~ and \limit—is pushed up before the \project~ operation.
  \item The physical plan after \patstrat~ is shown in \reffig{rbo-ic2}(c), which aligns with the user-specific order, namely \texttt{(p:PERSON) $\rightarrow$ (friend:PERSON) $\rightarrow$ (msg:POST|COMMENT)}.
  \item Subsequently, \gopt~ continues to optimize the query by applying \fusionrule~ and \dfrule~. The \fusionrule~is applied to merge the \expandedge~ and \getvertex~ operations into a single \expandvertex~ operation, reducing intermediate data pipeline overhead.
  \pkirule~is applied when a primary key indexing is available, enabling direct lookups instead of full scans, significantly reducing the amount of data processed. The physical plan after optimization is shown in \reffig{rbo-ic2}(d).
  \item Finally, \lprule~further enhances the physical plan, as illustrated in \reffig{rbo-ic2}(e). After \trimrule~application, the \code{(friend:PERSON)} node retains only the properties \texttt{id}, \texttt{firstName}, and \texttt{lastName}, while \code{(msg:POST|COMMENT)} node retains only the properties \texttt{id}, \texttt{creationDate}, \texttt{content}, and \texttt{imageFile}.
  However, \trimrule~does not determine when these properties should be retrieved—whether they should be prefetched at node/edge retrieval or lazily fetched at the final Projection.
  \lprule~ensures that properties are always retrieved lazily at projection time. This rule is particularly beneficial in single-node system scenario, where accessing local properties incurs no additional network overhead.
\end{itemize}

In order to further validate the efficiency of these \strategies, we conducted the ablation experiment to assess their individual contributions.
To mitigate the impact of \strategy~ dependencies on the ablation outcomes, 
We incrementally added the rules in their intended application order and measured the execution time after each addition. The results, as illustrated in \reftab{ic2_ablation_test}, provide insights into the effectiveness of each rule within the optimization process.

For most rules (excluding \filterrule~), the addition of each rule reduces the execution time by 15\% to 25\%, with an average reduction of approximately 22\%. While the individual impact of each rule is moderate, their cumulative effect significantly enhances performance, reducing the overall execution time from 371 ms to 134 ms—a speedup of nearly 3x.

Among all the rules, \filterrule~ stands out as the most impactful. When this rule is applied, the execution time drops dramatically from 10,599 ms to 371 ms, representing a 30x improvement in performance. This substantial gain is expected, as \filterrule~ performs critical data pruning at an early stage of execution, drastically reducing the number of nodes and edges that need to be processed in subsequent stages. This early filtering minimizes unnecessary computations, leading to a significant boost in efficiency.

\begin{table}
  \small
  \centering
  \caption{Ablation Results for $IC_2$.}
  \vspace*{-0.5em}
  \label{tab:ic2_ablation_test}
  \begin{tabular}{ l | r }
  \hline
    \bf{Rules} & \bf{Execution Time (ms)} \\
    \hline
    None & 10599 \\
    \hline
    +\filterrule~ & 371 \\
    \hline
    +\trimrule~ & 301 \\
    \hline
    +\sptrule~ & 251 \\
    \hline
    +\fusionrule~ & 208 \\
    \hline
    +\lprule~ & 156 \\
    \hline
    +\pkirule~ & 134 \\
    \hline
  \end{tabular}
\end{table}
  
\subsection{Multi-Patterns Heuristic Optimizaiton.}

\begin{figure*}[ht]
  \centering
  \includegraphics[scale=0.25]{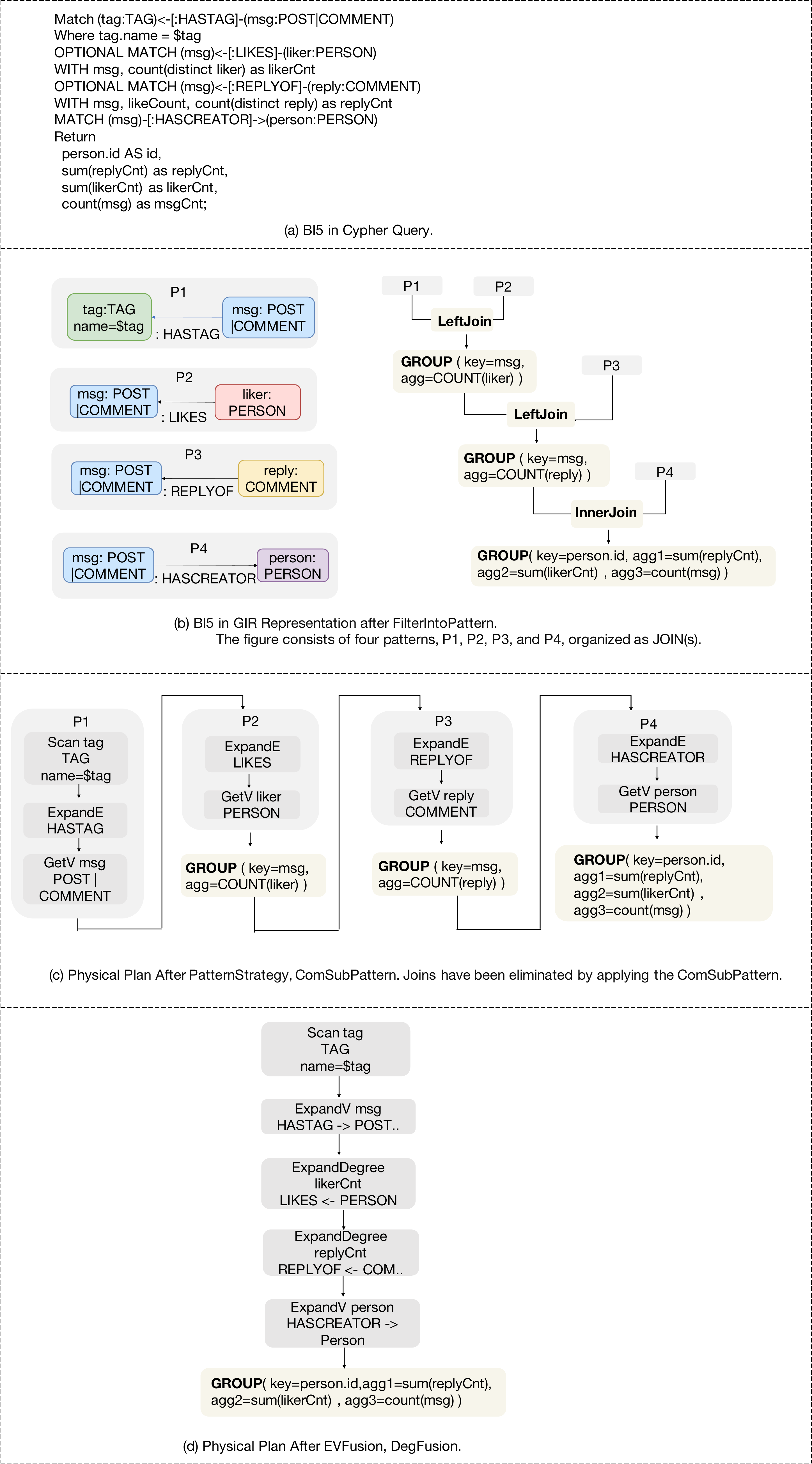}
  \caption{Optimization Process of $BI_5$.}
  \label{fig:rbo-bi5}
\end{figure*}

The query $BI_5$, shown in \reffig{rbo-bi5}(a), contains multiple MATCH clauses, some of which are optional, along with additional relational operations between these MATCH clauses.
The complexity of the query stems from the interaction between the MATCH and relational clauses. While each individual pattern is relatively simple, the overall query becomes significantly more complex.
To optimize such queries effectively, \gopt~ employs advanced optimization \strategies, such as \commonrule~, \dfrule~, and others, which are highlighted in blue in \reffig{bi5-dag}, and optimizes the query as follows:

\begin{figure}[H]
  \centering
  \includegraphics[width=1.0\linewidth]{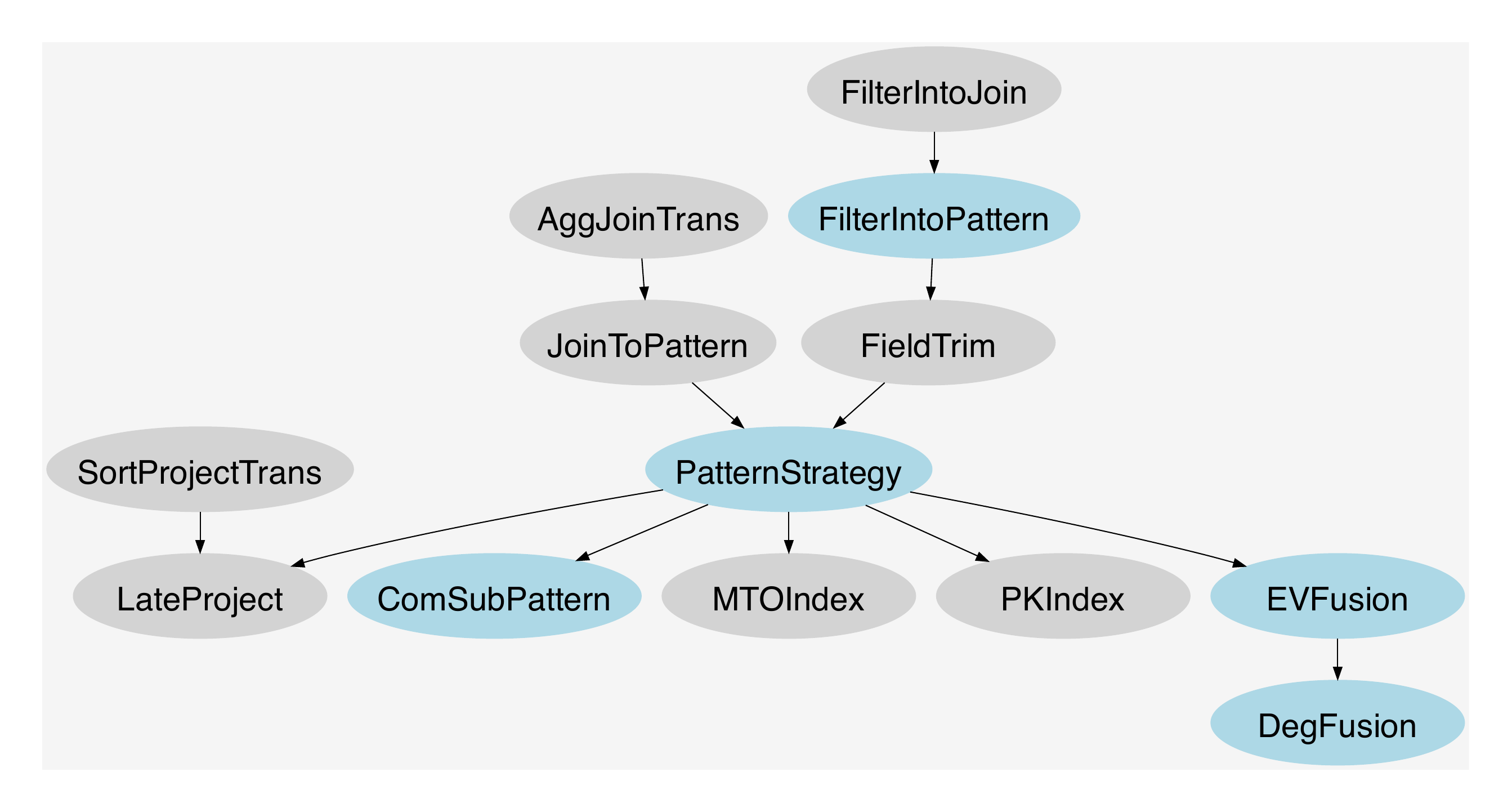}
  \caption{Strategies DAG of $BI_5$.}
  \label{fig:bi5-dag}
\end{figure}

\begin{itemize}
  \item \gopt~ first converts the query into a \ir~. The multiple MATCH clauses ($P_1$, $P_2$, $P_3$, $P_4$) are organized by \join~ operations. If the right MATCH is optional, the \join~ type is LEFT OUTER; otherwise, it is an INNER.
  The structure of the \join~ is followed by subsequent relational operations, which form the left branch of the next \join. The \ir~ representation after \filterrule~ is shown in \reffig{rbo-bi5}(b).
  \item Next, \gopt~ applies \patstrat~ for each pattern. The \patstrat~ preserves the user-defined order. For pattern $P_1$, the operations are output in the order of \texttt{(tag:TAG)} $\rightarrow$ \texttt{(msg:POST|COMMENT)}.
  The same optimization is applied to other patterns. Then, \commonrule~is applied, it identifies that $P_1$ and $P_2$ share the intermediate results of \texttt{(msg:POST|COMMENT)}, and since $P_1$ directly outputs the \texttt{msg} as the final result, the pattern operations in $P_2$ can be expanded from $P_1$’s output. As a result, the \join~ between $P_1$ and $P_2$ can be flattened into an expansion. The same reasoning applies to $P_2$, $P_3$, and $P_4$, resulting in the physical plan shown in \reffig{rbo-bi5}(c).
  \item Finally, \gopt~ applies two additional rules: \fusionrule~ and \dfrule~. The \fusionrule~fuses \expandedge~ and \getvertex~ into a single \expandvertex operation.
  This fusion enables the application of \dfrule~. The \dfrule~identifies that \getvertex~ is followed by a \group~ operation, where the \group~ key corresponds to the start vertex of \expandvertex, and the \group~ value is the distinctly-aggregated count of the target vertex of \expandvertex, which implies that the \group~ operation performs degree aggregation. Consequently, \expandvertex~ and \group~ can be merged into a single \expanddegree~ operation. The final physical plan is shown in \reffig{rbo-bi5}(d).
\end{itemize}

Additionally, we provide an ablation experiment to demonstrate the contribution of each rule in this optimization process. The results are shown in \reftab{bi5_ablation_test}.
The execution time decreases from 386 ms to 286 ms after incorporating both \dfrule~ and \fusionrule~, demonstrating that these two rules together can achieve a 25\% speedup. Upon adding \commonrule~, there is a significant reduction in execution time from 114,678 ms down to 386 ms, marking a remarkable 400x performance improvement. This substantial enhancement is anticipated because \commonrule~ optimizes the process by maximizing the reuse of intermediate results from previously identified common patterns and eliminating unnecessary \join~ operations, thus preventing superfluous data exploration. Additionally, \filterrule~ contributes to an approximate 3x performance improvement, as evidenced by the reduction in execution time from 400,414 ms to 114,678 ms upon its introduction.

\begin{table}
  \small
  \centering
  \caption{Ablation Results for $BI_5$.}
  \vspace*{-0.5em}
  \label{tab:bi5_ablation_test}
  \begin{tabular}{ l | r }
  \hline
    \bf{Rules} & \bf{Execution Time (ms)} \\
    \hline
    None   & 400414 \\
    \hline
    +\filterrule~ & 114678 \\
    \hline
    +\commonrule~ & 386 \\
    \hline
    +\fusionrule~ & 359 \\
    \hline
    +\dfrule~ & 286 \\
    \hline
  \end{tabular}
\end{table}

\subsection{Path Optimization.}

\begin{figure*}[ht]
  \centering
  \includegraphics[width=0.8\linewidth]{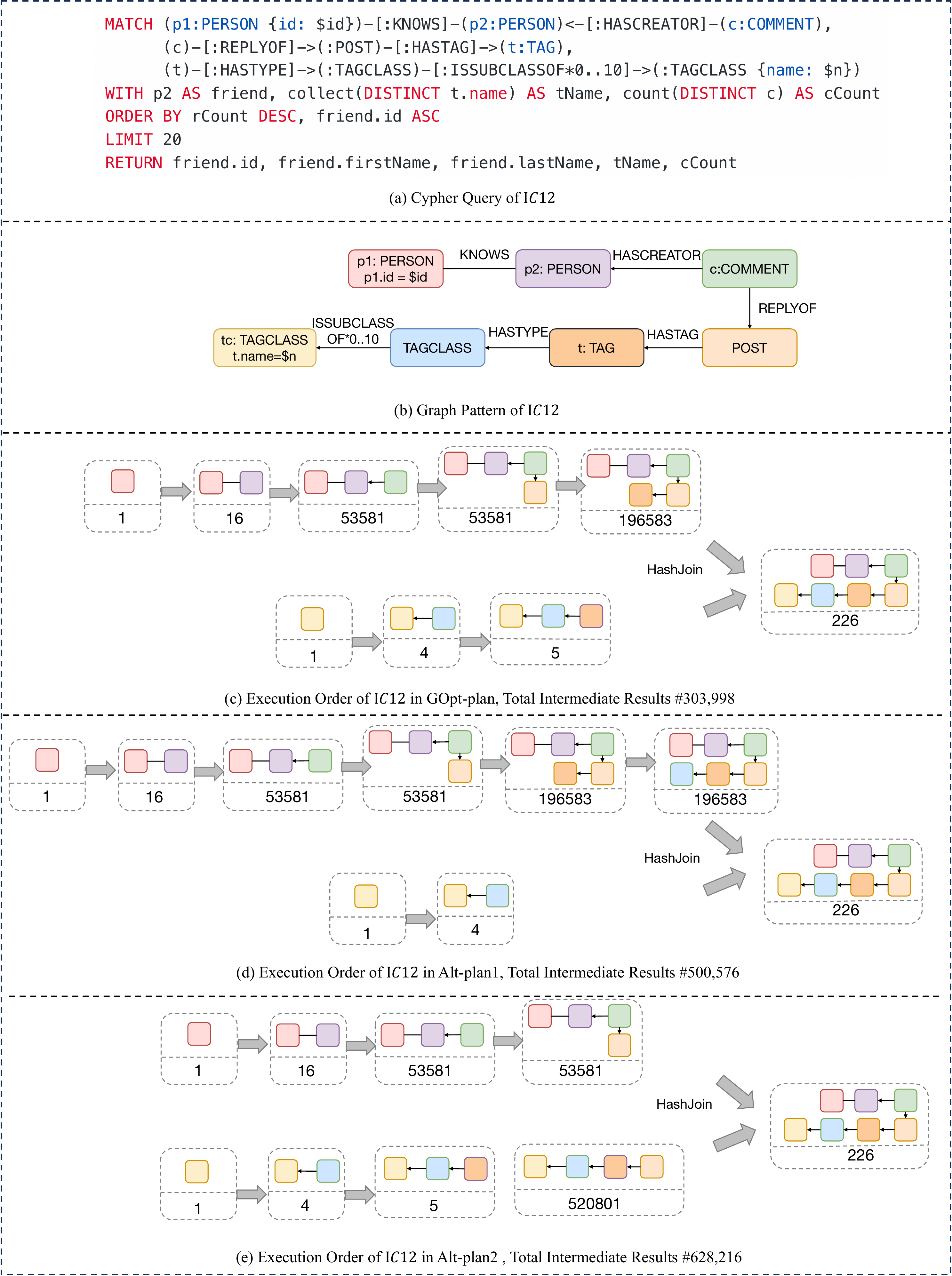}
  \caption{Case Study on $IC_{12}$}
  \label{fig:ic12}
\end{figure*}

\begin{table}
  \small
  \centering
  \caption{Time Costs for $IC_{12}$.}
  \vspace*{-0.5em}
  \label{tab:ic12_test}
  \begin{tabular}{ l | r }
  \hline
    \bf{Plan} & \bf{Execution Time (ms)} \\
    \hline
    \gopt~-plan & 175 \\
    \hline
    Alt-plan1 & 252 \\
    \hline
    Alt-plan2 & 438 \\
    \hline
  \end{tabular}
\end{table}

We illustrate the optimization of a path query $IC_{12}$ in \reffig{ic12}, with corresponding execution times for each plan shown in \reftab{ic12_test}. 
The query aims to find the path between two vertices: a \code{PERSON} $p_1$ with specified id, and a \code{TAGCLASS} $t$ with specified name. 
The Cypher query and the query pattern are shown in \reffig{ic12}(a) and \reffig{ic12}(b), respectively. 

The optimized plan generated by \\gopt~~is displayed in \reffig{ic12}(c), with the corresponding number of intermediate results marked at each expansion step. Similarly, the alternative plans in \reffig{ic12}(d) and \reffig{ic12}(e) are also annotated with the number of intermediate results.
In the optimized plan, the path is decomposed into two \bgps: $P_1$, starting from \code{PERSON} $p_1$, and $P_2$, starting from \code{TAGCLASS} $t$. The results of $P_1$ and $P_2$ are joined on the \code{TAG} vertex to produce the final results.
This optimized plan effectively reduces the number of intermediate results generated during query execution, thereby enhancing overall performance.
For comparison, we provide two alternative plans in \reffig{ic12}(d) and \reffig{ic12}(e), each joining on different vertices. 
These alternative plans are shown to be suboptimal as they generate more intermediate results compared to the optimized plan in \reffig{ic12}(c).

As demonstrated in \reftab{ic12_test}, the optimized plan generated by \\gopt~~achieves the lowest execution time of $175$ ms, outperforming the alternative plans by $30\%$ and $60\%$, respectively. This case study underscores the effectiveness of \\gopt~~in optimizing path queries.

\subsection{Cyclic Pattern Optimization.}

\begin{figure*}[ht]
  \centering
  \includegraphics[width=0.8\linewidth]{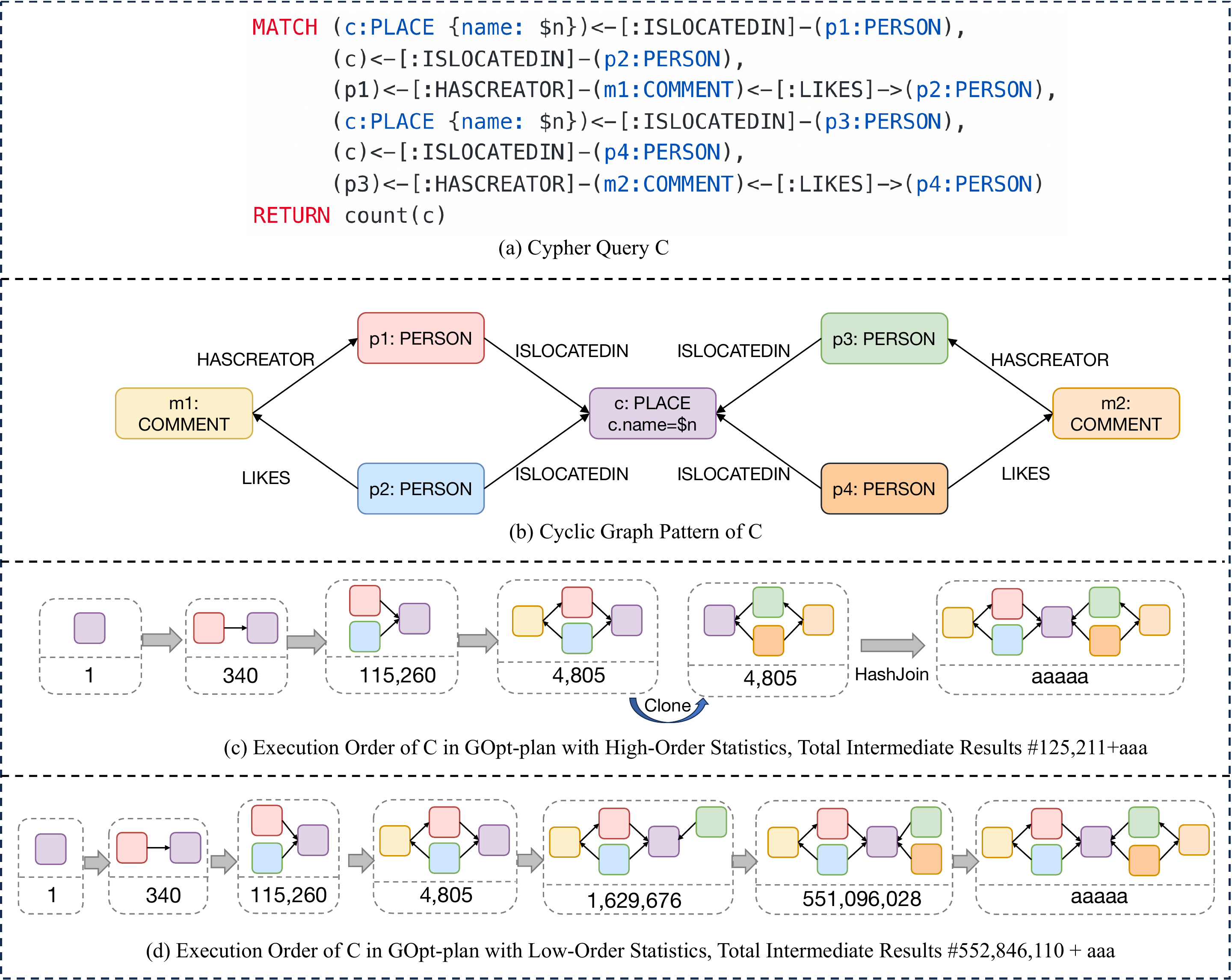}
  \caption{Case Study on Cyclic Query $C$}
  \label{fig:cyclic}
\end{figure*}

\begin{table}
  \small
  \centering
  \caption{Time Costs for Cyclic Query $C$.}
  \vspace*{-0.5em}
  \label{tab:cyclic_test}
  \begin{tabular}{ l | r }
  \hline
    \bf{Plan} & \bf{Execution Time (s)} \\
    \hline
    \gopt~-plan with High-Order Statistics & 1163 \\
    \hline
    \gopt~-plan with Low-Order Statistics & \ot (>3600) \\
    \hline
  \end{tabular}
\end{table}

In \reffig{cyclic}, we present a case study on optimizing a cyclic query $C$, with the corresponding execution times for each plan shown in \reftab{cyclic_test}.
The primary objective of this study is to determine the optimal search order for a cyclic pattern using cost-based optimization techniques. 
The Cypher query is illustrated in \reffig{cyclic}(a), with the corresponding query pattern shown in \reffig{cyclic}(b).

In \reffig{cyclic}(c) and \reffig{cyclic}(d), we present two optimized plans generated by \\gopt~, utilizing high-order and low-order statistics, respectively, and mark the number of intermediate results generated at each expansion step.
The comparison reveals that the optimized plan in \reffig{cyclic}(c) with high-order statistics is more efficient than the one in \reffig{cyclic}(d), as it generates fewer intermediate results during query execution. 
Specifically, although the search order in the first four expansion steps is identical for both plans, as the pattern expands and grows larger,
the optimized plan using low-order statistics produces significantly more intermediate results. This inefficiency arises from substantial deviations in the estimated cardinality compared to the actual values, leading to a suboptimal search order. 
In contrast, the high-order statistics enable more accurate cardinality estimations in the optimizer, resulting in a more efficient search order.
The execution times in \reftab{cyclic_test} further confirm the superiority of the optimized plan with high-order statistics, which completes the query in $1163$ s, while the plan with low-order statistics exceeds the timeout threshold of $3600$ s.
This case study demonstrates the importance of utilizing high-order statistics in cost-based optimization techniques to enhance query performance.

\subsection{Complex Pattern Optimization.}

\begin{figure*}[!ht]
  \centering
  \includegraphics[width=0.8\linewidth]{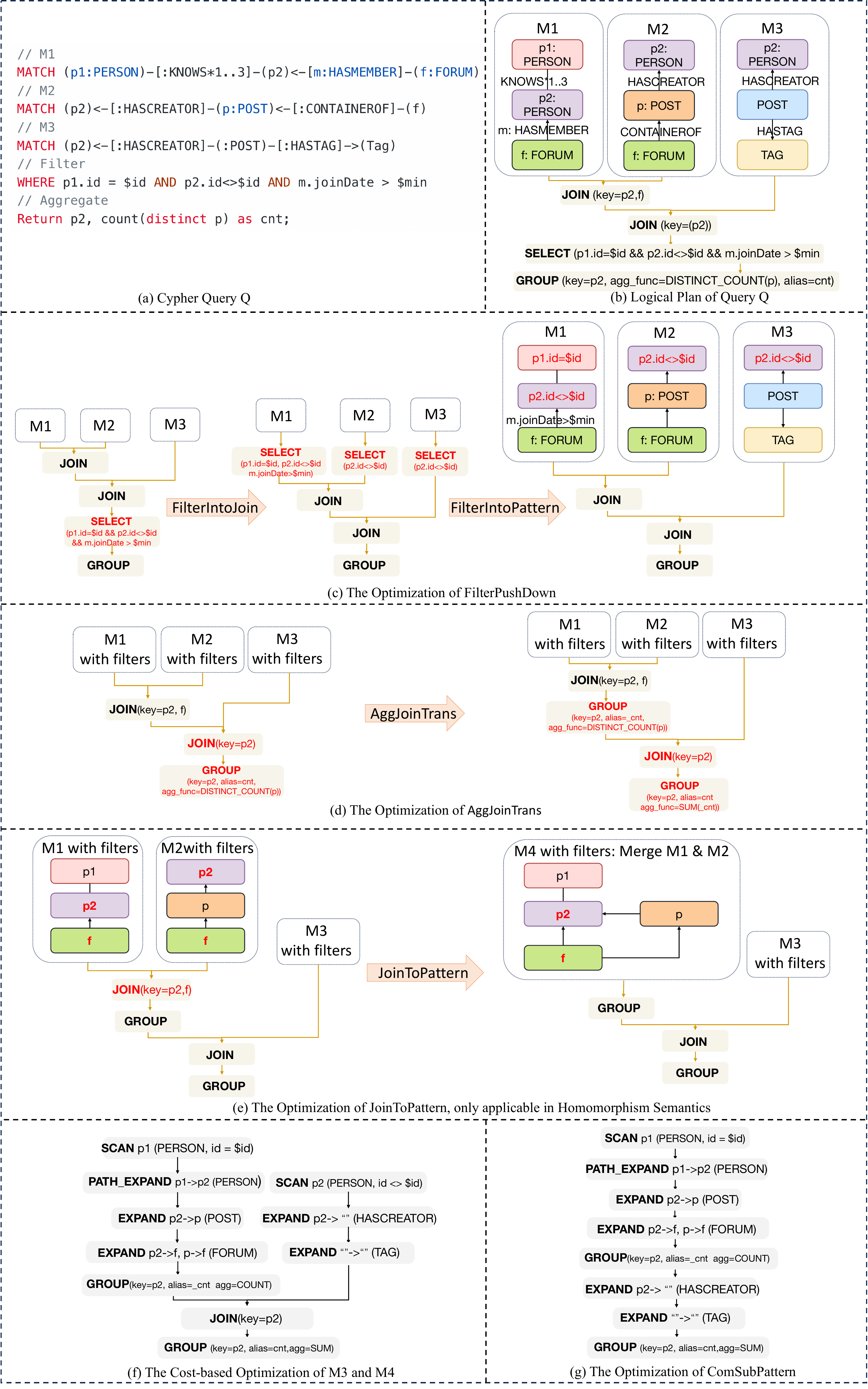}
  \caption{Case Study on Complex Query $Q$}
  \label{fig:ic56}
\end{figure*}

\begin{table}
  \small
  \centering
  \caption{Ablation Results for Complex Query $Q$.}
  \vspace*{-0.5em}
  \label{tab:ic56_test}
  \begin{tabular}{ l | r }
  \hline
    \bf{Applied Rule} & \bf{Execution Time (ms)} \\
    \hline
    None & 770573 \\
    \hline
    \filterrule~ & 234313 \\
    \hline
    \ajtrule~ & 214955 \\
    \hline
    \joinelimrule~ & 64466 \\
    \hline
    \commonrule~ & 7014 \\
    \hline
  \end{tabular}
\end{table}

We present a complex query $Q$, which combines $IC_5$ and $IC_6$, to demonstrate a comprehensive optimization process in \reffig{ic56}. This process involves both the \rulestrat~and \patstrat.

\stitle{\rulestrat.} 
The query pattern is depicted in \reffig{ic56}(a).
Initially, we outline the logical plan in \\gopt~~as shown in \reffig{ic56}(b), where the query is translated into three \bgps~$M_1$, $M_2$, and $M_3$, followed by relational operators \joinopr, \select, and \group.

The optimization process begins by attempting to push down the \select~operator. This technique reduces computational costs by minimizing the number of intermediate results. The optimized plan is presented in \reffig{ic56}(c). 
Here, we initially apply the existing relational rule \ftjrule~from Calcite, which integrates the \select~operator into the inputs of the join operations. 
Following this, we implement the \filterrule, which further advances the pushdown of the \select~operator into the \bgps~$M_1$, $M_2$ and $M_3$. 
By pushing the \select~operator to the \bgp~level, unnecessary graph elements are filtered out at earlier stages in the query processing, thereby enhancing performance.

Subsequently, we apply the \ajtrule~ to push an aggregate (i.e., \group) pass a \joinopr, which is also an existing relational rule in Calcite, as shown in \reffig{ic56}(d).
This rule focuses on pushing \group~operations further down the plan to limit intermediate results.
In our case, the \group~shares the same key as the \joinopr~(the one with the \code{key=p2}), which allows the \group~to progress through this join, with an additional \group~to accumulate the \code{\_cnt} attribute after the join, as the join operator may generate duplicate matches of \code{p2}. This is a standard optimization in relational query processing, reducing intermediate results prior to joining the \bgps~$M_3$.

Following these adjustments, we employ the \joinelimrule~to remove unnecessary join operators and merge the \bgps. It is important to note that the \joinelimrule~is applicable only under homomorphism semantics (as we have discussed in \refsec{semantics}), though for demonstration purposes, we apply this rule assuming such semantics, despite Cypher's use of edge-distinct semantics. As illustrated in \reffig{ic56}(e), \bgps~$M_1$ and $M_2$ are merged into a singular pattern, $M_{4}$, with the join keys \code{(p2,f)} as the common vertices in the merged pattern.

After pattern merging, we utilize cost-based optimization techniques to further refine the \bgps, seeking the most efficient query execution order. The resulting optimized order is specified in \reffig{ic56}(f). It is noteworthy that within the cyclic \bgp~$M_4$, we apply a \wcoj~strategy based on the backend \gs, leveraging \physicalspec, ensuring execution efficiency with worst-case optimality.

Finally, we apply the \commonrule, as shown in \reffig{ic56}(g), which further eliminates the join operator between $M_4$ followed by \group, and $M_3$. 
This rule can be applied since: (1) the \group~operation outputs \code{(p2,\_cnt)}, (2) the following join operator has join key \code{p2}, which is the subset of the output of the \group, and (3) the search order of \bgp~$M_3$ starts from the \code{p2}.
We eliminate the join operator and directly search the \bgp~$M_3$ by starting from the common vertices (i.e., matches of \code{p2}) in the output of the \group, to accelerate the query processing.

To validate the efficiency of these strategies further, we conducted an ablation experiment to assess their individual contributions, as shown in \reftab{ic56_test}. The results indicate that without any optimization rules, executing the query takes 770573 ms. Applying the \filterrule~achieves a 70\% improvement, reducing the execution time to 234313 ms. The \ajtrule~further enhances performance by 8\%. The \joinelimrule~achieves a 71\% improvement. Finally, the \commonrule~provides the most significant performance boost, achieving a 97\% improvement and reducing the execution time to 7014 ms. This case study demonstrates the effectiveness of \\gopt~~in optimizing complex queries through \rulestrat.
\begin{figure*}[ht]
  \centering
  \includegraphics[width=0.8\linewidth]{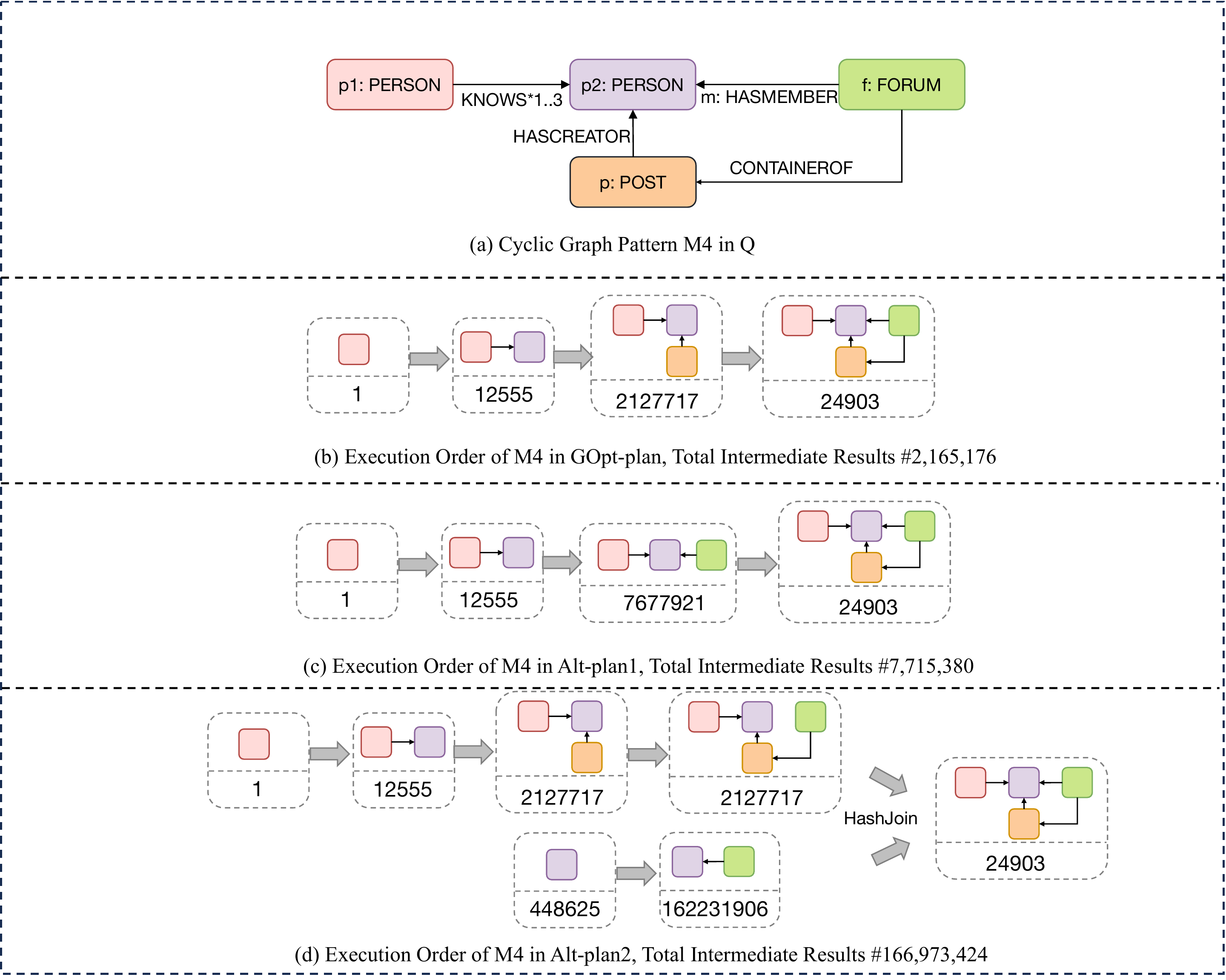}
  \caption{CBO on Cyclic Query $M4$}
  \label{fig:M4}
\end{figure*}

\begin{table}
  \small
  \centering
  \caption{Time Costs for $Q$, with different searching order of $M4$.}
  \vspace*{-0.5em}
  \label{tab:M4_test}
  \begin{tabular}{ l | r }
  \hline
    \bf{Plan} & \bf{Execution Time (ms)} \\
    \hline
    \gopt~-plan & 7014 \\
    \hline
    Alt-plan1 & 22211 \\
    \hline
    Alt-plan2 & 194047 \\
    \hline
  \end{tabular}
\end{table}
\stitle{\patstrat.} 
Beyond the overall \rulestrat~process discussed above, we delve into the detailed optimization process of \patstrat, particularly focusing on the cyclic pattern $M_4$ in the query $Q$.
The query pattern is illustrated in \reffig{M4}(a), while \reffig{M4}(b)-(d) present the optimized plan generated by \\gopt~ and two alternative plans, respectively. Each plan indicates the number of intermediate results produced at each expansion step.
The \\gopt~~optimized plan shown in \reffig{M4}(b) produces the fewest intermediate results compared to the alternatives in \reffig{M4}(c) and \reffig{M4}(d), resulting in a more efficient query execution. 

Additionally, \reftab{M4_test} provides execution times for $Q$ using different search orders of $M_4$.
These results confirm the superiority of the \\gopt~~optimized plan, which achieves the lowest execution time of $7014$ ms, outperforming the alternative plans by $3\times$ and $27\times$, respectively.
This case study underscores the effectiveness of \\gopt~~in optimizing complex queries with cyclic patterns, particularly in enhancing query performance through cost-based optimization techniques.
\section{Backend Integration}
\label{sec:converter}

The Physical Converter layer of \gopt~(see \refsec{architecture}) is responsible for transforming the optimized physical plan into a plan that is compatible with the underlying backend engine.
This layer serves as a bridge between \gopt~ and various backend engines, enabling seamless integration and execution of the optimized plans.

We offer two ways for integration.
The first one is through a \converter~interface, which defines the methods for converting each physical operator in the \gopt-optimized plan to the corresponding executable operator in the backend engine.
This allows \gopt~to traverse and convert the optimized physical plan into a backend-compatible executable plan.
We have integrated \gopt~with Neo4j using this approach.
The second one provides a Google Protocol Buffers (protobuf)~\cite{protobuf} based physical plan output to facilitate cross-platform integration.
This allows \gopt~to submit the optimized plan directly to the backend engine in protobuf format, enabling the backend engine to parse and transform it into its native execution plan.
We have integrated \gopt~with Alibaba's \gs~platform using this approach~\cite{graphscope_github}, enabling execution on the distributed dataflow engine \gaia~\cite{qian2021gaia}.

\subsection{\converter~Interface}

We define the \converter~interface as follows:
\begin{lstlisting}
    interface PhysicalConverter {
        // Convert methods for GIR's physical operators
        ExecOp convert(JOIN op);
        ExecOp convert(EXPAND op);
        ExecOp convert(PROJECT op);
        // Additional conversions for other physical operators are omitted for brevity
        ...
    }
\end{lstlisting}
Backends can implement the \converter~interface to convert each \girop~from the optimized physical plan into an \execop~that is executable by the backend engine.
For example, we implement the \converter~for Neo4j as follows:
\begin{lstlisting}
public class Neo4jConverter implements PhysicalConverter {
    // Context for logical planning in Neo4j
    private final LogicalPlanningContext context;
    // The logical plan producer in Neo4j,
    // which can construct the plan that is executable by Neo4j
    private final LogicalPlanProducer planProducer;
    // A mapper to transform GOpt's expressions to Neo4j's
    private final ExpressionMapper exprMapper;

    // Converts a JOIN into a Neo4j HashJoinPipe,
    // an executable plan node for Neo4j's query engine.
    @Override
    public HashJoinPipe convert(JOIN op) {
        // Transform the input plans into LogicalPlan nodes
        LogicalPlan left = op.getLeft().accept(this).getNode();
        LogicalPlan right = op.getRight().accept(this).getNode();
        // Convert the join condition to Neo4j's Variables
        Set<Variable> condition = toNeoVar(op.getCondition());
        // Use the plan producer to create a Neo4j HashJoinPipe
        return (HashJoinPipe) planProducer.planNodeHashJoin(condition, left, right, context);
    }


    // Converts a EXPAND into a Neo4j ExpandPipe,
    // an executable plan node for Neo4j's query engine.
    @Override
    public ExpandPipe convert(EXPAND op) {
        // Transform the input plan into a LogicalPlan node
        LogicalPlan input = op.getInput().accept(this).getNode();
        // The vertexExpansion can contain multiple edgeExpands
        List<EdgeExpand> edgeExpands = op.getExpands();
        // The first edgeExpand is converted into a
        // ExpandAll mode, expanding all the edges
        EdgeExpand first = edgeExpands.get(0);
        // Convert op fields to Neo4j structures
        String tag = op.getTag();
        String alias = op.getAlias();
        SemanticDirection dir = toNeoDir(op.getDir());
        List<RelType> relTypes = toNeoRelType(op.getRelTypes());
        Relationship rel = toNeoRel(tag, alias, dir, relTypes);
        // Define the ExpandAll mode for the first edgeExpand
        ExpansionMode expandAllMode = ExpandAll$.MODULE$;
        // Update the plan by adding the executable op
        ExpandPipe node = (ExpandPipe) planProducer.planSimpleExpand(input, tag, alias, rel, expandAllMode, context);
        // For the rest edgeExpands, we use the ExpandInto mode,
        // meaning that they are expanded and confirmed
        // to be existed in the first edgeExpand results
        for (int i = 1; i < edgeExpands.size(); i++) {
            EdgeExpand edgeExpand = edgeExpands.get(i);
            // Convert each field in edgeExpand, omit here
            ...
            // Use the ExpandInto mode
            ExpansionMode expandIntoMode = ExpandInto$.MODULE$;
            // Update the plan
            node = planProducer.planSimpleExpand(node, tag, alias, rel, expandIntoMode, context);
        }
        return node;
    }

    // Converts a PROJECT into a Neo4j ProjectPipe,
    // an executable plan node for Neo4j's query engine.
    @Override
    public ProjectPipe convert(PROJECT op) {
        // Transform the input plan into a LogicalPlan node
        LogicalPlan inputPlan = op.getInput().accept(this).getNode();
        // Transform the expression to Neo4j's expression
        List<Expression> exprs = op.getExprs().stream()
            .map(expr -> expr.accept(this))
            .collect(Collectors.toList());
        // Use the plan by adding the executable op
        return (ProjectPipe) planProducer.planRegularProjection(inputPlan, exprs, context);
    }

    // other conversion methods omitted for brevity ...
}
\end{lstlisting}

By implementing the \converter~interface, we observe that Neo4j's executable plan is incrementally constructed by the \texttt{LogicalPlanProducer} class (provided by Neo4j) through converting each physical operator in the \gopt-optimized plan in to the corresponding operator (named \code{XOpPipe}, which is a subclass of \code{ExecOp}) that is executable by Neo4j's query engine.

\subsection{Protobuf-based Physical Plan}
The second approach for backend integration utilizes a protobuf-based physical plan.
This design enables cross-platform integration and provides the complete view of the optimized physical plan to backend engines in a more flexible manner.
In this approach, \gopt~outputs the optimized physical plan in a serialized format using Google Protocol Buffers (protobuf) \cite{protobuf}, comprising all necessary information about the physical operators, their configurations, and the data flow between them.
By submitting the protobuf-based physical plan to the backend engines, these engines can parse the protobuf representation and transform it into their native execution plan format.
We have used this approach to integrate \gopt~with Alibaba's \gs~platform, facilitating execution on the distributed dataflow engine \gaia~\cite{qian2021gaia}.
We show an example as follows.

Firstly, in \gopt, we provide a build-in implementation of the \converter~interface that serializes the physical plan into a protobuf-based representation:
\begin{lstlisting}
    public class ProtobufConverter implements PhysicalConverter {
        // The protobuf builder for the physical plan
        private final PhysicalPlanPB.Builder planBuilder;

        // Convert methods for protobuf-based physical operators
        @Override
        public ExecOp convert(EXPAND op) {
            // Convert EXPAND to protobuf representation
            ExpandPB expandPB = ExpandPB.newBuilder()
                .addAllExpands(op.getExpands().stream()
                    .map(expand -> transformEdgeExpandPb(expand))
                    .collect(Collectors.toList()))
                .build();
            planBuilder.addOps(expandPB);
            return op;
        }

        // Additional conversions for other physical operators are omitted for brevity
        ...
    }
\end{lstlisting}

Then we submit the protobuf-based physical plan to the integrated backend engine \gaia~ in \gs:
\begin{lstlisting}
    // Serialize the physical plan to protobuf format
    PhysicalPlanPB physicalPlan = planBuilder.build();
    byte[] serializedPlan = physicalPlan.toByteArray();
    // Submit the serialized plan to the backend engine
    GaiaClient gaiaClient = new GaiaClient();
    gaiaClient.submit(serializedPlan);
\end{lstlisting}

In the engine side, \gaia~(which is developed in Rust) receives the serialized protobuf-based physical plan and further construct its native execution plan:

\begin{lstlisting}
    // GAIA is a dataflow-based engine for distributed graph processing
    worker.dataflow(move |input, output| {
        // Deserialize the protobuf-based physical plan
        let physicalPlan = PhysicalPlanPb::decode(serializedPlan);
        // Retrieve source data from the graph based on the Source operator
        let sources = get_source_data(graph, physicalPlan.getSource());
        let mut stream = input.input_from(sources);
        // Iterate through the protobuf operators and convert them to native operators
        for op : physicalPlan.getOpsList() {
            let op_kind = op.getOpKind();
            match op_kind {
                OpKind::Expand(expand) -> {
                    // parse the pb operator and generate the udf
                    let func = gen_expand_udf(edge)?;
                    // add a flat_map operator with the udf
                    stream = stream.flat_map(move |input| func.exec(input))?;
                }
                OpKind::Project(project) -> {
                    // parse the pb operator and generate the udf
                    let func = gen_project_udf(project)?;
                    // add a map operator with the udf
                    stream = stream.map(move |input| func.exec(input))?;
                }
                // Additional cases for other physical operators are omitted for brevity
                ...
            }
        }
        // Finally, sink the stream into the output
        stream.sink_into(output)?;
    });
\end{lstlisting}

In this way, the protobuf-based physical plan offers a flexible and efficient approach for \gaia~integration, allowing \gopt~to seamlessly interact with cross-platform backend engines.

\section{Conclusion}
\label{sec:conlusion}

In this paper, we extends the SIGMOD 2025 paper "A Modular Graph-Native Query Optimization Framework" by offering an in-depth exploration of \gopt's advanced technical mechanisms, implementation \strategies, and extended evaluations. Our focus lies on detailing the optimization \strategies~ that \gopt utilizes to enhance the efficiency of complex graph queries, specifically highlighting \rulestrat~ and \patstrat~ methodologies. Through the utilization of query cases derived from the LDBC Social Network Benchmark (SNB), we illustrate the potent capabilities of these combined \strategies~ in optimizing such queries. Empirical experiments conducted further elucidate the individual and collective contributions of each \strategy~ towards overall optimization efficacy. Finally, we introduce the physical converter layer in \gopt, showcasing its role in facilitating seamless integration with different backend engines.

\bibliographystyle{ACM-Reference-Format}
\bibliography{main}


\begin{thebibliography}{18}


\ifx \showCODEN    \undefined \def \showCODEN     #1{\unskip}     \fi
\ifx \showDOI      \undefined \def \showDOI       #1{#1}\fi
\ifx \showISBNx    \undefined \def \showISBNx     #1{\unskip}     \fi
\ifx \showISBNxiii \undefined \def \showISBNxiii  #1{\unskip}     \fi
\ifx \showISSN     \undefined \def \showISSN      #1{\unskip}     \fi
\ifx \showLCCN     \undefined \def \showLCCN      #1{\unskip}     \fi
\ifx \shownote     \undefined \def \shownote      #1{#1}          \fi
\ifx \showarticletitle \undefined \def \showarticletitle #1{#1}   \fi
\ifx \showURL      \undefined \def \showURL       {\relax}        \fi
\providecommand\bibfield[2]{#2}
\providecommand\bibinfo[2]{#2}
\providecommand\natexlab[1]{#1}
\providecommand\showeprint[2][]{arXiv:#2}

\bibitem[ant(2024)]%
        {antlr}
 \bibinfo{year}{2024}\natexlab{}.
\newblock \bibinfo{title}{ANTLR: ANother Tool for Language Recognition}.
\newblock
\newblock
\urldef\tempurl%
\url{https://www.antlr.org/}
\showURL{%
\tempurl}


\bibitem[gra(2024)]%
        {graphscope_github}
 \bibinfo{year}{2024}\natexlab{}.
\newblock \bibinfo{title}{GraphScope's open-source project on GitHub}.
\newblock
\newblock
\urldef\tempurl%
\url{https://github.com/alibaba/GraphScope}
\showURL{%
\tempurl}


\bibitem[neo(2024)]%
        {neo4j}
 \bibinfo{year}{2024}\natexlab{}.
\newblock \bibinfo{title}{Neo4j Graph Database}.
\newblock \bibinfo{howpublished}{https://neo4j.com/}.
\newblock


\bibitem[ajt(2025)]%
        {ajtrule}
 \bibinfo{year}{2025}\natexlab{}.
\newblock \bibinfo{title}{AggJoinTrans in Calcite}.
\newblock
\newblock
\urldef\tempurl%
\url{https://calcite.apache.org/javadocAggregate/org/apache/calcite/rel/rules/AggregateJoinTransposeRule.html}
\showURL{%
\tempurl}


\bibitem[tri(2025)]%
        {trimrule}
 \bibinfo{year}{2025}\natexlab{}.
\newblock \bibinfo{title}{FieldTrimRule in Calcite}.
\newblock
\newblock
\urldef\tempurl%
\url{https://calcite.apache.org/javadocAggregate/org/apache/calcite/sql2rel/RelFieldTrimmer.html}
\showURL{%
\tempurl}


\bibitem[ftj(2025)]%
        {ftjrule}
 \bibinfo{year}{2025}\natexlab{}.
\newblock \bibinfo{title}{FilterIntoJoin in Calcite}.
\newblock
\newblock
\urldef\tempurl%
\url{https://calcite.apache.org/javadocAggregate/org/apache/calcite/rel/rules/FilterJoinRule.html}
\showURL{%
\tempurl}


\bibitem[spt(2025)]%
        {sptrule}
 \bibinfo{year}{2025}\natexlab{}.
\newblock \bibinfo{title}{SortProjectTrans in Calcite}.
\newblock
\newblock
\urldef\tempurl%
\url{https://calcite.apache.org/javadocAggregate/org/apache/calcite/rel/rules/SortProjectTransposeRule.html}
\showURL{%
\tempurl}


\bibitem[sup(2025)]%
        {supported_grammars}
 \bibinfo{year}{2025}\natexlab{}.
\newblock \bibinfo{title}{Supported Grammars in GOpt}.
\newblock
\newblock
\urldef\tempurl%
\url{https://github.com/alibaba/GraphScope/tree/main/interactive_engine/compiler/src/main/antlr4}
\showURL{%
\tempurl}


\bibitem[Angles et~al\mbox{.}(2017)]%
        {angles17}
\bibfield{author}{\bibinfo{person}{Renzo Angles}, \bibinfo{person}{Marcelo
  Arenas}, \bibinfo{person}{Pablo Barcel\'{o}}, \bibinfo{person}{Aidan Hogan},
  \bibinfo{person}{Juan Reutter}, {and} \bibinfo{person}{Domagoj Vrgo\v{c}}.}
  \bibinfo{year}{2017}\natexlab{}.
\newblock \showarticletitle{Foundations of Modern Query Languages for Graph
  Databases}.
\newblock \bibinfo{journal}{\emph{ACM Comput. Surv.}} \bibinfo{volume}{50},
  \bibinfo{number}{5}, Article \bibinfo{articleno}{68} (\bibinfo{date}{sep}
  \bibinfo{year}{2017}), \bibinfo{numpages}{40}~pages.
\newblock
\showISSN{0360-0300}
\urldef\tempurl%
\url{https://doi.org/10.1145/3104031}
\showDOI{\tempurl}


\bibitem[Begoli et~al\mbox{.}(2018)]%
        {Begoli_2018}
\bibfield{author}{\bibinfo{person}{Edmon Begoli}, \bibinfo{person}{Jesús
  Camacho-Rodríguez}, \bibinfo{person}{Julian Hyde},
  \bibinfo{person}{Michael~J. Mior}, {and} \bibinfo{person}{Daniel Lemire}.}
  \bibinfo{year}{2018}\natexlab{}.
\newblock \showarticletitle{Apache Calcite: A Foundational Framework for
  Optimized Query Processing Over Heterogeneous Data Sources}. In
  \bibinfo{booktitle}{\emph{Proceedings of the 2018 International Conference on
  Management of Data}} \emph{(\bibinfo{series}{SIGMOD/PODS ’18})}.
  \bibinfo{publisher}{ACM}.
\newblock
\urldef\tempurl%
\url{https://doi.org/10.1145/3183713.3190662}
\showDOI{\tempurl}


\bibitem[Fan et~al\mbox{.}(2021)]%
        {graphscope}
\bibfield{author}{\bibinfo{person}{Wenfei Fan}, \bibinfo{person}{Tao He},
  \bibinfo{person}{Longbin Lai}, \bibinfo{person}{Xue Li},
  \bibinfo{person}{Yong Li}, \bibinfo{person}{Zhao Li},
  \bibinfo{person}{Zhengping Qian}, \bibinfo{person}{Chao Tian},
  \bibinfo{person}{Lei Wang}, \bibinfo{person}{Jingbo Xu},
  \bibinfo{person}{Youyang Yao}, \bibinfo{person}{Qiang Yin},
  \bibinfo{person}{Wenyuan Yu}, \bibinfo{person}{Jingren Zhou},
  \bibinfo{person}{Diwen Zhu}, {and} \bibinfo{person}{Rong Zhu}.}
  \bibinfo{year}{2021}\natexlab{}.
\newblock \showarticletitle{GraphScope: A Unified Engine for Big Graph
  Processing}.
\newblock \bibinfo{journal}{\emph{Proc. VLDB Endow.}} \bibinfo{volume}{14},
  \bibinfo{number}{12} (\bibinfo{date}{jul} \bibinfo{year}{2021}),
  \bibinfo{pages}{2879–2892}.
\newblock


\bibitem[Francis et~al\mbox{.}(2018)]%
        {cypher}
\bibfield{author}{\bibinfo{person}{Nadime Francis}, \bibinfo{person}{Alastair
  Green}, \bibinfo{person}{Paolo Guagliardo}, \bibinfo{person}{Leonid Libkin},
  \bibinfo{person}{Tobias Lindaaker}, \bibinfo{person}{Victor Marsault},
  \bibinfo{person}{Stefan Plantikow}, \bibinfo{person}{Mats Rydberg},
  \bibinfo{person}{Petra Selmer}, {and} \bibinfo{person}{Andr{\'e}s Taylor}.}
  \bibinfo{year}{2018}\natexlab{}.
\newblock \showarticletitle{Cypher: An evolving query language for property
  graphs}. In \bibinfo{booktitle}{\emph{Proceedings of the 2018 International
  Conference on Management of Data}}. \bibinfo{pages}{1433--1445}.
\newblock


\bibitem[{LDBC Social Network Benchmark}(2022)]%
        {ldbc_snb}
\bibfield{author}{\bibinfo{person}{{LDBC Social Network Benchmark}}.}
  \bibinfo{year}{2022}\natexlab{}.
\newblock \showarticletitle{\url{https://ldbcouncil.org/benchmarks/snb/}}.
  \bibinfo{publisher}{[Online; accessed 20-October-2022]}.
\newblock


\bibitem[Lyu et~al\mbox{.}(2024)]%
        {lyu2024gopt}
\bibfield{author}{\bibinfo{person}{Bingqing Lyu}, \bibinfo{person}{Xiaoli
  Zhou}, \bibinfo{person}{Longbin Lai}, \bibinfo{person}{Yufan Yang},
  \bibinfo{person}{Yunkai Lou}, \bibinfo{person}{Wenyuan Yu}, {and}
  \bibinfo{person}{Jingren Zhou}.} \bibinfo{year}{2024}\natexlab{}.
\newblock \bibinfo{title}{A Modular Graph-Native Query Optimization Framework}.
\newblock
\newblock
\showeprint[arxiv]{2401.17786}~[cs.DB]
\urldef\tempurl%
\url{https://arxiv.org/abs/2401.17786}
\showURL{%
\tempurl}


\bibitem[{Protocol Buffers}(2024)]%
        {protobuf}
\bibfield{author}{\bibinfo{person}{{Protocol Buffers}}.}
  \bibinfo{year}{2024}\natexlab{}.
\newblock \showarticletitle{\url{https://protobuf.dev/overview/}}.
\newblock


\bibitem[Qian et~al\mbox{.}(2021)]%
        {qian2021gaia}
\bibfield{author}{\bibinfo{person}{Zhengping Qian}, \bibinfo{person}{Chenqiang
  Min}, \bibinfo{person}{Longbin Lai}, \bibinfo{person}{Yong Fang},
  \bibinfo{person}{Gaofeng Li}, \bibinfo{person}{Youyang Yao},
  \bibinfo{person}{Bingqing Lyu}, \bibinfo{person}{Xiaoli Zhou},
  \bibinfo{person}{Zhimin Chen}, {and} \bibinfo{person}{Jingren Zhou}.}
  \bibinfo{year}{2021}\natexlab{}.
\newblock \showarticletitle{{GAIA}: A System for Interactive Analysis on
  Distributed Graphs Using a {High-Level} Language}. In
  \bibinfo{booktitle}{\emph{18th USENIX Symposium on Networked Systems Design
  and Implementation (NSDI 21)}}. \bibinfo{publisher}{USENIX Association},
  \bibinfo{pages}{321--335}.
\newblock
\showISBNx{978-1-939133-21-2}
\urldef\tempurl%
\url{https://www.usenix.org/conference/nsdi21/presentation/qian-zhengping}
\showURL{%
\tempurl}


\bibitem[Rodriguez(2015)]%
        {gremlin}
\bibfield{author}{\bibinfo{person}{Marko~A. Rodriguez}.}
  \bibinfo{year}{2015}\natexlab{}.
\newblock \showarticletitle{The Gremlin Graph Traversal Machine and Language
  (Invited Talk)}. In \bibinfo{booktitle}{\emph{Proceedings of the 15th
  Symposium on Database Programming Languages}} (Pittsburgh, PA, USA)
  \emph{(\bibinfo{series}{DBPL 2015})}. \bibinfo{publisher}{Association for
  Computing Machinery}, \bibinfo{address}{New York, NY, USA},
  \bibinfo{pages}{1–10}.
\newblock
\showISBNx{9781450339025}
\urldef\tempurl%
\url{https://doi.org/10.1145/2815072.2815073}
\showDOI{\tempurl}


\bibitem[Sz{\'a}rnyas et~al\mbox{.}(2018)]%
        {Szrnyas2018ReducingPG}
\bibfield{author}{\bibinfo{person}{G{\'a}bor Sz{\'a}rnyas},
  \bibinfo{person}{J{\'o}zsef Marton}, \bibinfo{person}{J{\'a}nos Maginecz},
  {and} \bibinfo{person}{D{\'a}niel Varr{\'o}}.}
  \bibinfo{year}{2018}\natexlab{}.
\newblock \showarticletitle{Reducing Property Graph Queries to Relational
  Algebra for Incremental View Maintenance}.
\newblock \bibinfo{journal}{\emph{ArXiv}}  \bibinfo{volume}{abs/1806.07344}
  (\bibinfo{year}{2018}).
\newblock
\urldef\tempurl%
\url{https://api.semanticscholar.org/CorpusID:49313167}
\showURL{%
\tempurl}


\end{thebibliography}
\begin{appendix}
    \section{APPENDIX}

\label{sec:appendix}

\end{appendix}
\end{document}